\documentclass[12pt]{article}

\usepackage{authblk}
\usepackage[top=25truemm,bottom=28truemm,left=24truemm,right=24truemm]{geometry}
\usepackage{setspace}
\normalsize
\usepackage{amsmath,amssymb,amsthm,mathtools,bm,mathrsfs}
\usepackage{physics}
\usepackage{braket}
\usepackage{graphicx}
\usepackage{color}
\usepackage{fancyhdr}
\usepackage{cite}
\usepackage{xurl}
\usepackage[dvipdfmx,colorlinks=true,linkcolor=blue,citecolor=blue,urlcolor=blue]{hyperref}
\numberwithin{equation}{section}

\newcommand{\AdS}{\mathrm{AdS}}
\newcommand{\CFT}{\mathrm{CFT}}
\renewcommand{\dd}{\mathrm{d}}

\newcommand{\Ecal}{\mathcal{E}}
\newcommand{\Ocal}{\mathcal{O}}

\renewcommand{\headrulewidth}{0pt}
\title{\bf Boundary Energy Density and Null Propagation in Global AdS\(_3\)/CFT\(_2\)}
\author[]{Shiki Yoshikawa}
\affil[]{\it Department of Physics, Kyoto University, Kyoto 606-8502, Japan\\
{\small\tt yoshikawa.shiki@gauge.scphys.kyoto-u.ac.jp}}
\date{\today}

\begin{document}

\maketitle
\thispagestyle{fancy}
\renewcommand{\headrulewidth}{0pt}

\begin{abstract}

We study how bulk propagation in global
AdS$_3$ is reflected in boundary energy densities of the dual
CFT$_2$ on the cylinder.  A key feature of the global geometry,
which is absent in the Poincar\'e-patch description, is that a null
excitation emitted from the boundary reaches the antipodal boundary
point after \(\Delta\tau=\pi\) and returns to the original boundary
point after \(\Delta\tau=2\pi\).  We show that the CFT energy density
reproduces this sequence through two chiral peaks that move along the
boundary light-cone directions, meet at the antipodal point, and return
to the original point after one global period.
For a wave-packet state, the leading energy density consists of two
chiral peaks moving along the boundary light-cone directions; their
relative weights encode the impact parameter of the corresponding bulk
null ray.
For a state constructed from an operator of large conformal
dimension with a Euclidean time regulator (LCD state), the
exact cylinder stress tensor gives two periodic chiral pulses moving
along the boundary light-cone directions. These pulses reach the
antipodal point and return to the original point at the same global
times as the radial null geodesic in the bulk.
Thus the boundary energy density provides a
time-dependent boundary profile that reflects characteristic global
features of null propagation in AdS$_3$.

\end{abstract}

\newpage
\thispagestyle{empty}
\tableofcontents

\newpage
\pagestyle{plain}

\section{Introduction and summary}

One of the basic questions in holography is how bulk propagation is
encoded in the dual boundary theory in AdS/CFT
\cite{Maldacena:1997re,Gubser:1998bc,Witten:1998qj}. More generally,
this question is part of the broader problem of bulk reconstruction and
the emergence of local bulk physics in AdS/CFT
\cite{Hamilton:2006fh,Hamilton:2006az,Kabat:2011rz,Harlow:2018fse}.
In this paper we focus on a simple aspect of this problem: how
boundary-to-boundary bulk propagation in global \(\AdS_3\) is reflected
in boundary energy densities of the dual \(\CFT_2\) on the cylinder.

The global patch is particularly useful for this purpose. A null ray
emitted from the boundary reaches the antipodal boundary point after
\(\Delta\tau=\pi\) and returns to the original boundary point after
\(\Delta\tau=2\pi\). This boundary-to-boundary propagation is not visible
in the same way in a Poincar\'e-patch description. The natural question is
therefore how this global propagation appears in boundary observables
on the cylinder. This is closely related to the appearance of
bulk-cone singularities in Lorentzian CFT correlators, which reflect
bulk null propagation \cite{Hubeny:2006yu,Dodelson:2023nnr}.

In this paper we focus on the boundary energy density as a simple
observable that can be computed explicitly in the states we consider.
We will not assume that the energy density literally draws the bulk
trajectory. Rather, we ask whether it reflects the arrival of the bulk
null ray at the antipodal boundary point, its subsequent return to the
original boundary point, and the associated periodicity of global AdS
propagation.\footnote{We do not assume that the bulk motion must be
reproduced as a literal local propagation on the boundary
\cite{Tanahashi:2025fqi}. The point is rather that the
boundary-to-boundary propagation and periodicity implied by bulk motion
should be encoded consistently in boundary observables. See also
\cite{Terashima:2020tub} for related discussions on bulk locality and
bulk local states.}
The energy density is a time-dependent one-point observable in the
excited state and remains nonzero also at intermediate times, while the
wave packet propagates through the AdS interior.  Thus it describes the
evolution of a physical quantity throughout the bulk propagation, in
the same sense as the Poincar\'e-patch energy density studied in
\cite{Terashima:2023mcr}, rather than only detecting the singularity of
a boundary two-point function.
\footnote{
This is different from probes of bulk-cone singularities, where bulk null
propagation is inferred from singularities of boundary correlation
functions when the insertion points are connected by a bulk null
geodesic \cite{Hubeny:2006yu}.
}

Wave packets have played an important role in studies of bulk locality
and scattering in AdS/CFT
\cite{Gary:2009ae,Gary:2009mi}. A useful starting point for the present
problem is the wave-packet construction of \cite{Terashima:2023mcr},
where the CFT energy density associated with a bulk wave packet in
\(\AdS_3/\CFT_2\) was shown to be localized in two light-like peaks.
Since that analysis was carried out in a Poincar\'e patch, it does not
directly display the periodic boundary-to-boundary propagation of global
AdS. We therefore reformulate the wave-packet analysis on the global
cylinder and study how the local light-like peaks appear as periodic
chiral peaks in the boundary energy density.
The wave-packet construction superposes many global normal modes while
selecting narrow relative ranges of energy and angular momentum, thereby
producing a localized state that admits a semiclassical WKB description.
Although the underlying stress-tensor correlator is fixed by conformal
symmetry, smearing the boundary operator produces a state-dependent
boundary energy density that can be compared with the propagation of the
corresponding bulk wave packet.  In particular, the two chiral weights
reproduce the bulk energy and angular momentum through their sum and
difference, while their relative asymmetry encodes the impact parameter
of the bulk null ray.

We also study a second class of states constructed from a local
scalar primary operator of large conformal dimension, with a Euclidean
time regulator. We refer to this state as the large-conformal-dimension
operator state (LCD state). This construction is
related to earlier holographic local-quench setups, where local CFT
excitations were used to probe localized bulk dynamics
\cite{Nozaki:2013wia,Caputa:2014vaa}, and more directly to the localized global-AdS states constructed in \cite{Berenstein:2019qmm}. More generally, Euclidean sources provide a
way to prepare Lorentzian semiclassical bulk states in holographic CFTs
\cite{Marolf:2017kvq}. This example is useful for our purpose because
the boundary stress tensor can be computed exactly. The resulting energy
density consists of two chiral pulses moving along the boundary
light-cone directions. The energy density remains nonzero
throughout the evolution.  In the small-regulator regime, the
global-mode analysis in Appendix \ref{app:heavy-global-modes} shows that
the peak of the corresponding bulk wave function approaches the radial
null trajectory. We compare this behavior with the corresponding
bulk profile and show that the boundary pulses reach the antipodal point
and return to the original point at the same global times as the radial
null geodesic in global AdS. Thus this state provides another example in which the periodic boundary-to-boundary propagation of a localized bulk excitation is
visible in the boundary energy density.

In Section \ref{sec:null-wave}, we first review null geodesics in
global \(\AdS_3\) and then analyze the boundary wave-packet state. We
derive the energy density from exact cylinder correlators. In
the small-width regime, the leading contribution comes from local
neighborhoods of the nearest pole. The resulting
energy density consists of two periodic chiral peaks moving along the
boundary light-cone directions. 
Here ``chiral'' refers to the holomorphic and anti-holomorphic
components of the two-dimensional CFT stress tensor. In Lorentzian
signature these components propagate along the boundary light-cone
directions
\[
u=\frac{t}{L}+\theta,
\qquad
v=\frac{t}{L}-\theta .
\]
Their relative weights encode the impact
parameter of the corresponding bulk null ray, while their propagation
on the cylinder reproduces the arrival at the antipodal boundary point
and the subsequent return to the original boundary point.

In Section \ref{sec:heavy-null}, we analyze the LCD state. We compute the
exact cylinder stress tensor and find two periodic chiral pulses in the
boundary energy density. We then compare this result with the bulk
profile in global AdS. The boundary
pulses reach the antipodal point and return to the original point at the
same global times as this radial null propagation.

In Section \ref{sec:discussion}, we summarize the comparison between
the two examples and discuss possible extensions. In particular, we
comment on how the periodic structure found in pure global AdS may be
modified in black-hole backgrounds, and on which features of the result
are special to \(\AdS_3/\CFT_2\). 

The appendices provide the bulk interpretation of the two states used
in the main text. In Appendix \ref{app:bulk-wave-packet}, we analyze the
bulk behavior of the wave-packet state using global Klein--Gordon
modes. In Appendix \ref{app:heavy-global-modes}, we study the bulk
profile of the LCD state using
global AdS modes.

The main conclusion of this paper is that the global cylinder
description makes the boundary-to-boundary propagation and periodicity
of localized bulk excitations visible in boundary energy densities. For the
wave-packet state, the leading energy density is obtained
from the exact cylinder correlators. It remains nonzero while the
bulk wave packet propagates in the AdS interior, and the two chiral
components meet at the antipodal point after \(\Delta\tau=\pi\).
Their integrated weights reproduce the bulk energy and angular
momentum, while their relative weights encode the impact parameter.
For the LCD state, the exact boundary stress tensor shows the same
arrival at the antipodal boundary point and subsequent return to the
original point as the radial null geodesic. Together, these examples
show that boundary energy densities on the cylinder encode the
boundary-to-boundary propagation and periodicity of localized bulk
excitations in global \(\AdS_3\).

\section{Null geodesics and wave-packet states}
\label{sec:null-wave}

\subsection{Null geodesics in global \texorpdfstring{$\AdS_3$}{AdS3}}

In this subsection we use the dimensionless global time $\tau$. The corresponding dimensionful boundary time used in the CFT calculation below is $t=L\tau$. We use the global $\AdS_3$ metric
\begin{equation}
\dd s^2 = \frac{L^2}{\cos^2\rho}\left(-\dd\tau^2+\dd\rho^2+\sin^2\rho\,\dd\theta^2\right),
\qquad
0\leq \rho<\frac{\pi}{2},
\qquad
\theta\sim \theta+2\pi .
\label{eq:adsmetric}
\end{equation}
The conformal boundary is the cylinder $\mathbb{R}\times S^1$ with metric
\begin{equation}
\dd s^2_{\partial}\sim -\dd\tau^2+\dd\theta^2 .
\end{equation}

We first consider general null geodesics. Because $\tau$ and $\theta$ are Killing coordinates, the motion has two conserved quantities,
\begin{equation}
E:=-p_\tau=\frac{L^2}{\cos^2\rho}\dot\tau,
\qquad
J:=p_\theta=\frac{L^2\sin^2\rho}{\cos^2\rho}\dot\theta=L^2\tan^2\rho\,\dot\theta,
\label{eq:nullcharges}
\end{equation}
where the dot denotes differentiation with respect to an affine parameter. It is convenient to introduce the impact parameter\footnote{Strictly speaking, in global AdS this quantity is simply the
ratio of the conserved angular momentum and energy along the geodesic.
We nevertheless refer to it as an impact parameter by analogy with the
standard terminology in geodesic motion; see also \cite{Jia:2026pmv}
for a related discussion.}
\begin{equation}
b:=\frac{J}{E} .
\end{equation}
The null condition $\dd s^2=0$ gives
\begin{equation}
-\dot\tau^2+\dot\rho^2+\sin^2\rho\,\dot\theta^2=0
\end{equation}
after removing the overall conformal factor in \eqref{eq:adsmetric}. Using \eqref{eq:nullcharges}, we obtain
\begin{equation}
\dot\rho^2=\frac{\cos^4\rho}{L^4}\left(E^2-\frac{J^2}{\sin^2\rho}\right).
\end{equation}
Dividing by $\dot\tau^2$, this becomes
\begin{equation}
\left(\frac{\dd\rho}{\dd\tau}\right)^2=1-\frac{b^2}{\sin^2\rho} .
\label{eq:null-rho}
\end{equation}
Hence the turning point $\rho=\rho_*$ is determined by
\begin{equation}
\sin\rho_* = |b|,
\end{equation}
so a boundary-to-boundary null geodesic exists for $0\leq |b|\leq 1$. The case $|b|=1$ is the limiting boundary null curve.

The angular motion is obtained from
\begin{equation}
\frac{\dd\theta}{\dd\tau}=\frac{\dot\theta}{\dot\tau}=\frac{J}{E\sin^2\rho}=\frac{b}{\sin^2\rho} .
\end{equation}
Choosing $\tau=0$ at a boundary point and following the geodesic into the bulk, one finds from \eqref{eq:null-rho}
\begin{equation}
\cos\rho(\tau)=\sqrt{1-b^2}\,\sin\tau,
\qquad
0\leq \tau\leq \pi .
\label{eq:nullsol}
\end{equation}
At $\tau=0$ and $\tau=\pi$ one has $\rho=\pi/2$, while at $\tau=\pi/2$ one reaches the turning point $\rho=\rho_*$. Using the angular equation together with \eqref{eq:nullsol}, we obtain
\begin{equation}
\frac{\dd\theta}{\dd\tau}
=\frac{b}{1-(1-b^2)\sin^2\tau}
=\frac{b}{\cos^2\tau+b^2\sin^2\tau} .
\end{equation}
For $b>0$, this integrates to
\begin{equation}
\theta(\tau)=\theta_0+
\begin{cases}
\arctan(b\tan\tau), & 0\leq \tau\leq \pi/2,\\[0.4em]
\pi+\arctan(b\tan\tau), & \pi/2\leq \tau\leq \pi .
\end{cases}
\label{eq:theta-null}
\end{equation}
For $b<0$ the angular orientation is reversed. In either case, modulo $2\pi$,
\begin{equation}
\theta(0)=\theta_0,
\qquad
\theta(\pi)=\theta_0+\pi .
\end{equation}
Thus the boundary endpoints are antipodal independently of $b$, although the radial depth of the geodesic depends on $b$.

\begin{figure}[t]
  \centering
  \includegraphics[width=0.35\textwidth]{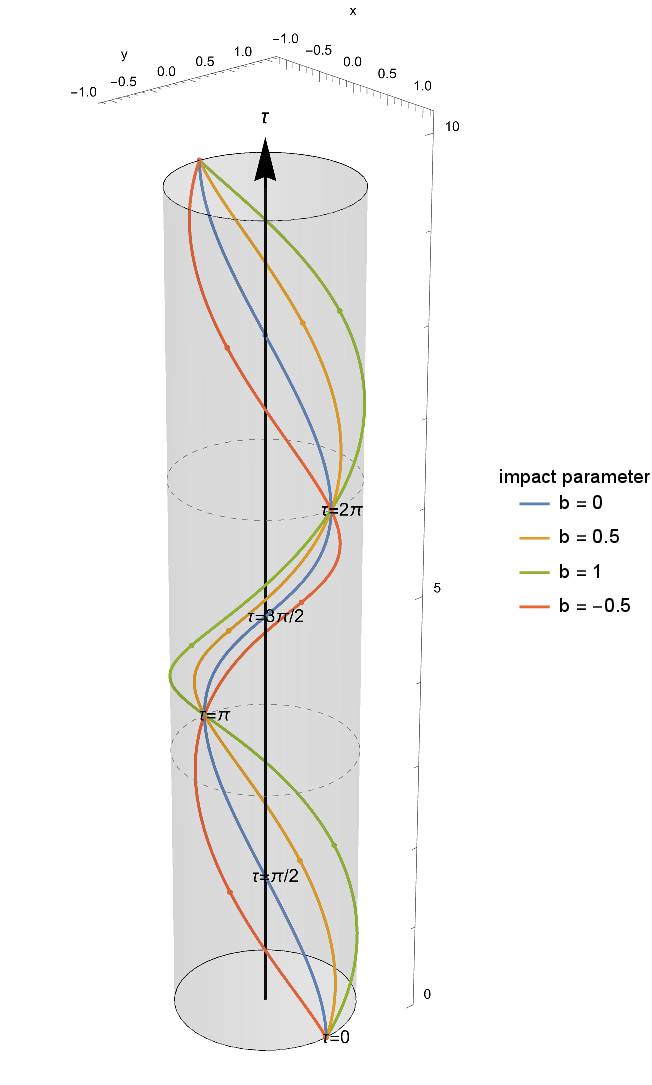}
  \caption{
  Bulk null geodesics in global AdS\(_3\) shown in a compactified spatial disk times the global time direction, for several values of the impact parameter \(b\).
  The geodesics start from the boundary at \(\tau=0\), reach a turning point at \(\tau=\pi/2\), arrive at the antipodal boundary point at \(\tau=\pi\), and return to the original boundary point at \(\tau=2\pi\).
  Different values of \(b\) change the depth of the trajectory in the bulk, while the boundary-to-boundary data \(\Delta \tau=\pi\) and \(\Delta\theta=\pi\) are unchanged. The curve with \(|b|=1\) should be understood as the limiting boundary null curve.
  }
  \label{fig:null_geodesics_bulk}
\end{figure}

Therefore the boundary-to-boundary data for null geodesics are
\begin{equation}
\Delta\tau^{\rm null}_{\rm bulk}=\pi,
\qquad
\Delta\theta^{\rm null}_{\rm bulk}=\pi.
\end{equation}
In terms of the dimensionful boundary time $t=L\tau$, this is
\begin{equation}
\Delta t^{\rm null}_{\rm bulk}=\pi L .
\end{equation}
After one full global period the ray returns to the original boundary
point,
\begin{equation}
\Delta\tau_{\rm period}^{\rm null}=2\pi,
\qquad
\mbox{or equivalently}\qquad
\Delta t_{\rm period}^{\rm null}=2\pi L .
\end{equation}
This sequence of boundary arrivals is what we will compare with the
periodic chiral peaks in the boundary wave-packet energy density. This behavior is illustrated in Fig.~\ref{fig:null_geodesics_bulk}.

\subsection{Wave-packet state on the boundary cylinder}

We now switch to the dimensionful boundary time
\begin{equation}
t=L\tau .
\end{equation}
The boundary cylinder is therefore described by
\begin{equation}
(t,\theta)\in \mathbb{R}\times[-\pi,\pi),
\qquad
\theta\sim\theta+2\pi,
\end{equation}
with boundary metric proportional to $-\dd t^2+L^2\dd\theta^2$. We consider a scalar primary operator $\Ocal$ of dimension $\Delta$, so that
\begin{equation}
h=\bar h=\frac{\Delta}{2} .
\end{equation}
We define the smeared state with central frequency $\omega$ and central angular momentum $m_0$ by
\begin{equation}
|\omega,m_0\rangle
:=\int_{-\infty}^{\infty}\dd t\int_{-\pi}^{\pi}\dd\theta\,
 f_{\omega,m_0}(t,\theta)\,\Ocal(t,\theta)|0\rangle,
\label{eq:smeared-state}
\end{equation}
with
\begin{equation}
f_{\omega,m_0}(t,\theta)
:=\exp\left[-\frac{t^2+4L^2\tan^2(\theta/2)}{2a^2}- i\omega t+ i m_0\theta\right] .
\end{equation}
   A primary state by itself is not a localized semiclassical packet with sharply specified energy and angular momentum; the oscillatory Gaussian smearing constructs such a packet by superposing global modes within narrow bands around \(\omega L\) and \(m_0\). 
In the small-width regime, the angular part of the source is a Gaussian
smearing localized near \(\theta=0\), since
\[
2L\tan\frac{\theta}{2}\simeq L\theta .
\] This construction is motivated by the wave-packet state studied in \cite{Terashima:2023mcr}. The main difference is that here the smearing is placed on the global boundary cylinder. In this paper we work in the small-width and high-frequency wave-packet regime
\begin{equation}
a\ll L,
\qquad
\frac{(\omega L-m_0)a}{L}\gg 1,
\qquad
\frac{(\omega L+m_0)a}{L}\gg 1,
\qquad
\omega L>|m_0| .
\label{eq:main-regime}
\end{equation}
Equivalently, in terms of the light-cone momenta introduced below,
\begin{equation}
p_u=\omega L-m_0,
\qquad
p_v=\omega L+m_0,
\end{equation}
this regime is
\begin{equation}
a\ll L,
\qquad
\frac{p_u a}{L}\gg 1,
\qquad
\frac{p_v a}{L}\gg 1,
\qquad
p_u,p_v>0 .
\label{eq:lightcone-regime}
\end{equation}
The condition \(a\ll L\) means that the wave packet is localized on a
scale much smaller than the AdS radius. The additional conditions \(p_u a/L\gg1\) and \(p_v a/L\gg1\) ensure
that the resulting bulk wave packet is sharply peaked around the
central energy \(\omega\) and angular momentum \(m_0\), with the
direction of angular motion fixed by the sign of \(m_0\). In Appendix \ref{app:bulk-wave-packet}, we make
this bulk interpretation explicit using
global normal-mode expansion. The smearing prepares a bulk one-particle wave
packet sharply peaked around \(\Omega_0=\omega L\) and \(m=m_0\), and
the stationary-phase conditions of the bulk profile reproduce the null
geodesic equations in global \(\AdS_3\).

\subsection{Exact cylinder correlators}

To formulate the wave-packet analysis directly on the global cylinder, we use the exact cylinder correlators. In the small-width regime \(a\ll L\), the smearing integrals are dominated by local neighborhoods of the relevant light-cone poles, where the cylinder correlators reduce to the same local form used in the plane analysis of \cite{Terashima:2023mcr}.
Let
\begin{equation}
w=t_E+ i L\theta,
\qquad
\bar w=t_E- i L\theta
\end{equation}
be Euclidean cylinder coordinates, where $t_E$ has the dimension of length. The Euclidean two-point function is
\begin{equation}
\left\langle \Ocal(w_1,\bar w_1)\Ocal(w_2,\bar w_2)\right\rangle_E
=C_\Ocal
\left(\frac{1}{2L\sinh\frac{w_{12}}{2L}}\right)^\Delta
\left(\frac{1}{2L\sinh\frac{\bar w_{12}}{2L}}\right)^\Delta .
\end{equation}
After analytic continuation $t_E\to  i t$, we define
\begin{equation}
u_{ij}:=\frac{t_i-t_j+L(\theta_i-\theta_j)- i\epsilon_{ij}}{L},
\qquad
v_{ij}:=\frac{t_i-t_j-L(\theta_i-\theta_j)- i\epsilon_{ij}}{L},
\qquad
\epsilon_{ij}:=\epsilon_i-\epsilon_j,
\label{eq:uvij}
\end{equation}
with $\epsilon_1>\epsilon_2>\epsilon_3>0$. The $ i\epsilon$ prescription fixes the ordering of the operators; we follow the standard prescription reviewed in \cite{duffin_lecture}. Then
\begin{equation}
\left\langle \Ocal(t_1,\theta_1)\Ocal(t_2,\theta_2)\right\rangle
=C_\Ocal
\left(\frac{1}{2L i\sin(u_{12}/2)}\right)^\Delta
\left(\frac{1}{2L i\sin(v_{12}/2)}\right)^\Delta .
\end{equation}
Using the plane-to-cylinder map $z=e^{w/L}$, the stress-tensor transformation law, and the conformal Ward identity, one finds
\begin{equation}
\left\langle \Ocal_1 T_2 \Ocal_3\right\rangle
=F_u(1,2,3)\left\langle \Ocal_1\Ocal_3\right\rangle,
\end{equation}
where
\begin{equation}
F_u(1,2,3)
=-\frac{\Delta}{8L^2}
\left(\frac{\sin(u_{13}/2)}{\sin(u_{12}/2)\sin(u_{23}/2)}\right)^2
-\frac{c}{24L^2} .
\label{eq:Fu}
\end{equation}
The corresponding expression for the anti-holomorphic stress tensor is obtained by replacing $u$ with $v$. We denote it by $F_v(1,2,3)$, so that
\begin{equation}
\left\langle \Ocal_1\bar T_2 \Ocal_3\right\rangle
=F_v(1,2,3)\left\langle \Ocal_1\Ocal_3\right\rangle .
\end{equation}
The exact pole structure on the cylinder is manifest:
\begin{equation}
u_{12}=2\pi n_1,
\qquad
u_{23}=2\pi n_3,
\qquad
n_1,n_3\in\mathbb{Z},
\label{eq:pole-lattice-u}
\end{equation}
and similarly in the $v$ sector. The role of the image labels will be explained below.

\subsection{Energy density for wave-packet state}

The norm of the wave-packet state is
\begin{equation}
N_0:=\langle\omega,m_0|\omega,m_0\rangle
=\int \dd t_1\dd\theta_1\dd t_3\dd\theta_3\,
 f_{\omega,m_0}(t_1,\theta_1)^* f_{\omega,m_0}(t_3,\theta_3)
 \left\langle \Ocal_1\Ocal_3\right\rangle .
\end{equation}
The holomorphic numerator is
\begin{equation}
N_T(t_2,\theta_2)
:=\int \dd t_1\dd\theta_1\dd t_3\dd\theta_3\,
 f_{\omega,m_0}(t_1,\theta_1)^* f_{\omega,m_0}(t_3,\theta_3)
 \left\langle \Ocal_1T_2\Ocal_3\right\rangle,
\end{equation}
and similarly for $N_{\bar T}$.

With our convention for the cylinder stress tensor,
\begin{equation}
T^{\rm cyl}_{tt}=T+\bar T .
\end{equation}
We define the boundary energy density by subtracting the
vacuum contribution,
\begin{equation}
\Ecal(t_2,\theta_2)
:=\frac{1}{2\pi}
\left[\left\langle T_{tt}(t_2,\theta_2)\right\rangle_{\omega,m_0}
-\left\langle T_{tt}\right\rangle_0\right]
=\frac{1}{2\pi}\left[
\frac{N_T(t_2,\theta_2)+N_{\bar T}(t_2,\theta_2)}{N_0}
+\frac{c}{12L^2}
\right].
\label{eq:energy-def}
\end{equation}
The quantity \(\Ecal\) includes the factor \(1/(2\pi)\). With the dimensionful boundary time convention used here, the physical energy is
\begin{equation}
E^{\rm phys}_{\CFT}=L\int_{-\pi}^{\pi}\dd\theta\,\Ecal(t,\theta).
\end{equation}
Equivalently, the dimensionless cylinder energy is
\begin{equation}
\Omega_{\CFT}=L E^{\rm phys}_{\CFT} .
\end{equation}

\subsection{Small-width approximation}

For $a\ll L$, the smearing integrals are dominated by small neighborhoods of the relevant light-cone poles on the cylinder. In these neighborhoods, the cylinder correlators reduce to the same local functional form as in the plane analysis of \cite{Terashima:2023mcr}.

We introduce the pointwise light-cone variables
\begin{equation}
u\equiv \frac{t+L\theta}{L},
\qquad
v\equiv \frac{t-L\theta}{L},
\end{equation}
so that
\begin{equation}
t=\frac{L}{2}(u+v),
\qquad
\theta=\frac{u-v}{2},
\qquad
\dd t\,\dd\theta=\frac{L}{2}\dd u\,\dd v .
\end{equation}
For $a\ll L$, the smearing localizes the integral near $\theta=0$, and
\begin{equation}
4L^2\tan^2\left(\frac{\theta}{2}\right)\simeq L^2\theta^2 .
\end{equation}
Hence
\begin{equation}
-\frac{t^2}{2a^2}-\frac{L^2\theta^2}{2a^2}
=-\frac{L^2}{4a^2}(u^2+v^2),
\end{equation}
while the phase factor in $f_{\omega,m_0}(t_1,\theta_1)^*f_{\omega,m_0}(t_3,\theta_3)$ becomes
\begin{equation}
 i\omega t- i m_0\theta
=\frac{ i}{2}p_u u+\frac{ i}{2}p_v v,
\end{equation}
where
\begin{equation}
p_u:=\omega L-m_0,
\qquad
p_v:=\omega L+m_0 .
\label{eq:pu-pv}
\end{equation}
The condition $\omega L>|m_0|$ is equivalent to $p_u,p_v>0$.

Before taking the local approximation, let us clarify the relation
between the local computation below and the exact cylinder contour
integral. In the exact cylinder correlator, the observation-point
singularities form an infinite set of periodic images. For \(a\ll L\), the
Gaussian smearing makes the neighborhoods of different images well
separated. The local calculation below computes the residue near the
nearest image, and the periodic result is obtained by summing the
corresponding leading pole contributions from the periodic images.

We now compute explicitly the local contribution from the image poles
which dominates near \(u_2=0\).  This is the nearest image,
\(n_1=n_3=0\).  Near the corresponding observation-point singularities,
\(u_{12}\simeq0\) and \(u_{23}\simeq0\), the first term in
\eqref{eq:Fu} reduces to
\begin{equation}
-\frac{\Delta}{8L^2}
\left(\frac{\sin(u_{13}/2)}{\sin(u_{12}/2)\sin(u_{23}/2)}\right)^2
\simeq
-\frac{\Delta}{2L^2}\left(\frac{u_{13}}{u_{12}u_{23}}\right)^2 .
\label{eq:Fu-local}
\end{equation}
This gives the local contribution to the peak near \(u_2=0\).

Let us specify which image poles contribute at leading order near each peak.
The exact cylinder correlator has image poles
\begin{equation}
u_{12}=2\pi n_1,
\qquad
u_{23}=2\pi n_3,
\qquad
n_1,n_3\in\mathbb Z .
\end{equation}
For a peak centered at \(u_2\simeq2\pi k\), these pole conditions imply
\begin{equation}
u_1\simeq u_2+2\pi n_1,
\qquad
u_3\simeq u_2-2\pi n_3 .
\end{equation}
Since the smearing localizes the source insertions near
\(u_1\simeq0\) and \(u_3\simeq0\), the leading pole pair near this peak
is
\begin{equation}
n_1=-k,
\qquad
n_3=k .
\end{equation}
All other pole pairs force at least one of \(u_1\) or \(u_3\) to be
displaced from the smearing center by an amount of order \(2\pi\), and
are therefore exponentially suppressed as
\(\exp[-{\rm const.}\,L^2/a^2]\) for \(a\ll L\).  Thus the calculation
below is the local residue calculation for the nearest peak; the
periodic copies are obtained by shifting to the corresponding leading
pole pair and summing over \(k\).

The local expression for the holomorphic numerator is
\begin{align}
N_T(u_2,v_2)
&\simeq \left(\frac{L}{2}\right)^2\widetilde C_\Ocal
\int \dd u_1\dd v_1\dd u_3\dd v_3\,
\exp\left[-\frac{L^2}{4a^2}(u_1^2+v_1^2+u_3^2+v_3^2)\right]
\notag\\
&\quad\times
\exp\left[\frac{ i}{2}(p_u u_1+p_v v_1-p_u u_3-p_v v_3)\right]
\notag\\
&\quad\times
\left[-\frac{\Delta}{2L^2}\left(\frac{u_{13}}{u_{12}u_{23}}\right)^2-\frac{c}{24L^2}\right]
\frac{1}{(u_{13})^\Delta(v_{13})^\Delta} .
\label{eq:NT-local}
\end{align}
Here \(\widetilde C_\Ocal\) includes the constant factors from the local
limit of the two-point function, and
\begin{equation}
u_{ij}=u_i-u_j-\frac{ i\epsilon_{ij}}{L},
\qquad
v_{ij}=v_i-v_j-\frac{ i\epsilon_{ij}}{L},
\qquad
\epsilon_{ij}:=\epsilon_i-\epsilon_j,
\qquad
\epsilon_1>\epsilon_2>\epsilon_3>0 .
\end{equation}
The first term in the square brackets contains the two poles in the nearest image, \(u_1=u_2\) and
\(u_3=u_2\), and gives the leading local light-cone peak near \(u_2=0\).
The corresponding leading pole pairs near \(u_2=2\pi k\) give the
periodic copies of this peak on the cylinder.

\subsection{Contour integral and dominant pole contributions}

We now keep the intermediate residue analysis, since it explains which poles of the cylinder correlator produce the leading energy-density peaks. We follow the contour argument of \cite{Terashima:2023mcr,Tanahashi:2025fqi}, adapted to the present notation. In this subsection the explicit residue calculation is written for
\(\Delta\in\mathbb{Z}_{>0}\) for simplicity. For non-integer
\(\Delta\), the same contour deformation crosses branch cuts rather
than isolated poles. In that case, one can deform the contour to wrap
around the relevant branch cuts, as in \cite{Terashima:2023mcr}, and
the leading contribution is obtained from the discontinuity across the
cuts. Thus the residue language used below should be understood as a
convenient shorthand for the corresponding branch-cut contribution. The
final small-width expression is not restricted to integer operator
dimensions.

We first extract the first term in the square brackets of \eqref{eq:NT-local}, which contains double poles. Up to the overall prefactor
\begin{equation}
\left(\frac{L}{2}\right)^2\widetilde C_\Ocal\left(-\frac{\Delta}{2L^2}\right),
\end{equation}
we define
\begin{align}
A
&:=\int \dd u_1\dd v_1\dd u_3\dd v_3\,
\exp\left[-\frac{L^2}{4a^2}(u_1^2+v_1^2+u_3^2+v_3^2)
+\frac{ i}{2}(p_u u_1+p_v v_1-p_u u_3-p_v v_3)\right]
\notag\\
&\quad\times
\frac{u_{13}^2}{u_{12}^2u_{23}^2(u_{13})^\Delta(v_{13})^\Delta} .
\label{eq:Adef}
\end{align}
Completing the square gives
\begin{align}
A&=e^{-\frac{a^2}{2L^2}(p_u^2+p_v^2)}
\int \dd u_1\dd v_1\dd u_3\dd v_3
\exp\left[-\frac{L^2}{4a^2}\left(u_1-\frac{ i p_u a^2}{L^2}\right)^2
-\frac{L^2}{4a^2}\left(v_1-\frac{ i p_v a^2}{L^2}\right)^2\right]
\notag\\
&\quad\times
\exp\left[-\frac{L^2}{4a^2}\left(u_3+\frac{ i p_u a^2}{L^2}\right)^2
-\frac{L^2}{4a^2}\left(v_3+\frac{ i p_v a^2}{L^2}\right)^2\right]
\frac{u_{13}^2}{u_{12}^2u_{23}^2(u_{13})^\Delta(v_{13})^\Delta} .
\label{eq:Asquare}
\end{align}
The shifts are written in terms of the dimensionless variables $u,v$.

The $v$-dependent part is
\begin{align}
I_v
&:=e^{-\frac{a^2}{2L^2}p_v^2}
\int_{\mathbb{R}}\dd v_1\int_{\mathbb{R}}\dd v_3\,
\exp\left[-\frac{L^2}{4a^2}\left(v_1-\frac{ i p_v a^2}{L^2}\right)^2
-\frac{L^2}{4a^2}\left(v_3+\frac{ i p_v a^2}{L^2}\right)^2\right]
\frac{1}{\left(v_1-v_3-\frac{ i\epsilon_{13}}{L}\right)^\Delta} .
\label{eq:Ivdef}
\end{align}
Here $\epsilon_{13}=\epsilon_1-\epsilon_3>0$. Thus, for fixed real $v_3$, the singularity in the $v_1$ plane is located at
\begin{equation}
v_1=v_3+\frac{ i\epsilon_{13}}{L},
\end{equation}
slightly above the original $v_1$-integration contour.

For integer $\Delta$, we use
\begin{equation}
\frac{1}{\left(v_1-v_3-\frac{ i\epsilon_{13}}{L}\right)^\Delta}
=\frac{(-1)^{\Delta-1}}{\Gamma(\Delta)}\partial_{v_1}^{\Delta-1}
\frac{1}{v_1-v_3-\frac{ i\epsilon_{13}}{L}} .
\end{equation}
After integrating by parts $\Delta-1$ times, this gives
\begin{align}
I_v
&=e^{-\frac{a^2}{2L^2}p_v^2}
\frac{1}{\Gamma(\Delta)}
\int_{\mathbb{R}}\dd v_1\int_{\mathbb{R}}\dd v_3\,
\frac{1}{v_1-v_3-\frac{ i\epsilon_{13}}{L}}
\notag\\
&\quad\times
\partial_{v_1}^{\Delta-1}
\exp\left[-\frac{L^2}{4a^2}\left(v_1-\frac{ i p_v a^2}{L^2}\right)^2
-\frac{L^2}{4a^2}\left(v_3+\frac{ i p_v a^2}{L^2}\right)^2\right] .
\end{align}
For $p_v>0$, the saddle of the $v_1$ Gaussian is at
\begin{equation}
v_1=\frac{ i p_v a^2}{L^2} .
\end{equation}
We deform the $v_1$-integration contour upward to
\begin{equation}
v_1\in\mathbb{R}+\frac{ i p_v a^2}{L^2} .
\end{equation}
Because of the $ i\epsilon$ prescription, the pole $v_1=v_3+ i\epsilon_{13}/L$ lies between the original contour and the deformed one. Therefore the deformation crosses this pole. The integral over the deformed contour is exponentially suppressed in the wave-packet regime $p_v a/L\gg 1$, and the leading contribution comes from the crossed pole.

Taking the residue, we obtain
\begin{align}
I_v
&\simeq \frac{2\pi i}{\Gamma(\Delta)}e^{-\frac{a^2}{2L^2}p_v^2}
\int_{\mathbb{R}}\dd v_3\,
\partial_{v_1}^{\Delta-1}
\exp\left[-\frac{L^2}{4a^2}\left(v_1-\frac{ i p_v a^2}{L^2}\right)^2
-\frac{L^2}{4a^2}\left(v_3+\frac{ i p_v a^2}{L^2}\right)^2\right]_{v_1=v_3} .
\end{align}
In the regime $p_v a/L\gg 1$, the derivative is dominated by
\begin{equation}
\left.\partial_{v_1}^{\Delta-1}
\exp\left[-\frac{L^2}{4a^2}\left(v_1-\frac{ i p_v a^2}{L^2}\right)^2\right]\right|_{v_1=v_3}
\simeq
\left(\frac{ i p_v}{2}\right)^{\Delta-1}
\exp\left[-\frac{L^2}{4a^2}\left(v_3-\frac{ i p_v a^2}{L^2}\right)^2\right] .
\end{equation}
The exponential factors combine as
\begin{align}
&e^{-\frac{a^2}{2L^2}p_v^2}
\exp\left[-\frac{L^2}{4a^2}\left(v_3-\frac{ i p_v a^2}{L^2}\right)^2
-\frac{L^2}{4a^2}\left(v_3+\frac{ i p_v a^2}{L^2}\right)^2\right]
=\exp\left[-\frac{L^2v_3^2}{2a^2}\right] .
\end{align}
Thus
\begin{equation}
I_v\simeq (2\pi)^{3/2}\frac{a}{L}\frac{ i^\Delta}{\Gamma(\Delta)}\left(\frac{p_v}{2}\right)^{\Delta-1},
\label{eq:Iv-result}
\end{equation}
up to corrections suppressed in the large $p_va/L$ expansion.

Next we consider the $u$-dependent part,
\begin{align}
I_u
&:=e^{-\frac{a^2}{2L^2}p_u^2}
\int\dd u_1\dd u_3\,
\exp\left[-\frac{L^2}{4a^2}\left(u_1-\frac{ i p_u a^2}{L^2}\right)^2
-\frac{L^2}{4a^2}\left(u_3+\frac{ i p_u a^2}{L^2}\right)^2\right]
\frac{u_{13}^2}{u_{12}^2u_{23}^2(u_{13})^\Delta} .
\label{eq:Iudef}
\end{align}
Here
\begin{equation}
u_{12}=u_1-u_2-\frac{ i\epsilon_{12}}{L},
\qquad
u_{23}=u_2-u_3-\frac{ i\epsilon_{23}}{L},
\qquad
\epsilon_{12},\epsilon_{23}>0 .
\end{equation}
Thus the pole $u_{12}=0$ lies slightly above the original $u_1$ contour, while the pole $u_{23}=0$ lies slightly below the original $u_3$ contour. For $p_u>0$, the saddle of the $u_1$ Gaussian lies at $u_1= i p_u a^2/L^2$. We deform the $u_1$ contour upward to this saddle. Because of the $ i\epsilon$ prescription, this deformation crosses the pole $u_{12}=0$. The integral over the deformed contour is exponentially suppressed in the wave-packet regime $p_u a/L\gg 1$, and the leading contribution comes from this crossed pole.

Taking the double-pole residue at $u_1=u_2$, we obtain
\begin{align}
I_u
&\simeq 2\pi i\,e^{-\frac{a^2}{2L^2}p_u^2}
\exp\left[-\frac{L^2}{4a^2}\left(u_2-\frac{ i p_u a^2}{L^2}\right)^2\right]
\int\dd u_3\,
\exp\left[-\frac{L^2}{4a^2}\left(u_3+\frac{ i p_u a^2}{L^2}\right)^2\right]
\notag\\
&\quad\times
\frac{1}{(u_{23})^\Delta}
\left[\frac{ i p_u}{2}-\frac{L^2u_2}{2a^2}-(\Delta-2)\frac{1}{u_{23}}\right] .
\label{eq:Iu-after-u1}
\end{align}
The term proportional to $u_2$ is suppressed by $(p_u a/L)^{-1}$ near the leading peak and will be omitted below. 

We next deform the $u_3$ contour downward to the saddle $u_3=- i p_u a^2/L^2$. Because of the $ i\epsilon$ prescription, this deformation crosses the pole $u_{23}=0$. The integral over the deformed contour is again exponentially suppressed, so the leading contribution comes from the crossed pole at $u_3=u_2$. Evaluating the residue gives
\begin{equation}
I_u\simeq
-\frac{8\pi^2}{\Delta\Gamma(\Delta)} i^\Delta
\left(\frac{p_u}{2}\right)^\Delta
\exp\left[-\frac{L^2u_2^2}{2a^2}\right],
\label{eq:Iu-result}
\end{equation}
up to corrections suppressed in the large $p_u a/L$ expansion.

Finally, let us comment on another singularity at $u_1=u_3$, which comes from the two-point factor $\langle\Ocal_1\Ocal_3\rangle$. This singularity is distinct from the observation-point poles $u_1=u_2$ and $u_3=u_2$. If one sets $u_1=u_3$ in the Gaussian part of the shifted integrand, including the prefactor from completing the square, one obtains
\begin{align}
&e^{-\frac{a^2p_u^2}{2L^2}}
\exp\left[-\frac{L^2}{4a^2}\left(u_3-\frac{ i p_u a^2}{L^2}\right)^2
-\frac{L^2}{4a^2}\left(u_3+\frac{ i p_u a^2}{L^2}\right)^2\right]
=\exp\left[-\frac{L^2u_3^2}{2a^2}\right] .
\end{align}
Thus this contribution is not exponentially suppressed by $e^{-a^2p_u^2/(2L^2)}$. Rather, it is subleading because, unlike the contribution obtained by
taking residues at both \(u_{12}=0\) and \(u_{23}=0\), it does not
receive the same large-\(p_u a/L\) enhancement. The leading light-cone peak is therefore produced by the pair of residues
at \(u_1=u_2\) and \(u_3=u_2\), while the singularity at \(u_1=u_3\)
gives only subleading corrections in the wave-packet regime.

For comparison, the two-point normalization integral in one light-cone sector is
\begin{align}
J(p)&:=e^{-\frac{a^2}{2L^2}p^2}
\int \dd u_1\dd u_3\,
\exp\left[-\frac{L^2}{4a^2}\left(u_1-\frac{ i p a^2}{L^2}\right)^2
-\frac{L^2}{4a^2}\left(u_3+\frac{ i p a^2}{L^2}\right)^2\right]
\frac{1}{(u_{13})^\Delta}
\notag\\
&\simeq (2\pi)^{3/2}\frac{a}{L}\frac{ i^\Delta}{\Gamma(\Delta)}\left(\frac{p}{2}\right)^{\Delta-1} .
\label{eq:Jp}
\end{align}
Thus the norm is
\begin{equation}
N_0\simeq \left(\frac{L}{2}\right)^2\widetilde C_\Ocal J(p_u)J(p_v) .
\end{equation}
The leading holomorphic numerator from the first term in $F_u(1,2,3)$ is
\begin{equation}
N_T^{\rm lead}(u_2)
\simeq \left(\frac{L}{2}\right)^2\widetilde C_\Ocal
\left(-\frac{\Delta}{2L^2}\right)I_uJ(p_v) .
\end{equation}
Using \eqref{eq:Iu-result} and \eqref{eq:Jp}, we find the leading contribution to the vacuum-subtracted stress-tensor component,
\begin{equation}
\frac{N_T^{\rm lead}(u_2)}{N_0}
\simeq \frac{\sqrt{2\pi}}{2aL}p_u
\exp\left[-\frac{L^2u_2^2}{2a^2}\right] .
\end{equation}
Because the energy density $\Ecal$ contains an additional factor $1/(2\pi)$, the leading holomorphic contribution is
\begin{equation}
\Ecal_u(t_2,\theta_2)
\simeq \frac{1}{2\sqrt{2\pi}\,aL}p_u
\exp\left[-\frac{L^2u_2^2}{2a^2}\right] .
\end{equation}
The anti-holomorphic contribution is obtained by the exchange
\begin{equation}
u\leftrightarrow v,
\qquad
p_u\leftrightarrow p_v,
\end{equation}
and gives
\begin{equation}
\Ecal_v(t_2,\theta_2)
\simeq \frac{1}{2\sqrt{2\pi}\,aL}p_v
\exp\left[-\frac{L^2v_2^2}{2a^2}\right] .
\end{equation}
Thus the local energy density is
\begin{equation}
\Ecal_{\rm local}(t_2,\theta_2)
\simeq \frac{1}{2\sqrt{2\pi}\,aL}
\left[p_u\exp\left(-\frac{L^2u_2^2}{2a^2}\right)
+p_v\exp\left(-\frac{L^2v_2^2}{2a^2}\right)\right],
\label{eq:E-local}
\end{equation}
where
\begin{equation}
u_2=\frac{t_2+L\theta_2}{L},
\qquad
v_2=\frac{t_2-L\theta_2}{L} .
\end{equation}
This is the local version of the two null peaks. With the convention above, this local expression carries
\begin{equation}
E^{\rm phys}_{\CFT}\simeq \frac{p_u+p_v}{2L}=\omega,
\end{equation}
before periodization, as expected from the central frequency of the state. Equivalently, the dimensionless cylinder energy is
\begin{equation}
\Omega_{\CFT}\simeq \frac{p_u+p_v}{2}=\omega L .
\end{equation}

\subsection{Periodization on the boundary cylinder}
Equation \eqref{eq:E-local} gives the contribution from the nearest
light-cone poles in the local patch.  Near every pole related by the
cylinder periodicity, the leading residue calculation takes the same
form after shifting \(u\) or \(v\) by an integer multiple of \(2\pi\).
Therefore, in the small-width regime, the energy density on the
cylinder is obtained by summing these leading image-pole contributions.

Under the angular identification \(\theta\sim\theta+2\pi\), one has
\begin{equation}
u\to u+2\pi,
\qquad
v\to v-2\pi,
\end{equation}
whereas the individual Gaussian factors in \eqref{eq:E-local} are not
periodic.  We therefore define the periodized Gaussian
\begin{equation}
S(x):=\sum_{n\in\mathbb{Z}}
\exp\left[-\frac{L^2(x-2\pi n)^2}{2a^2}\right],
\qquad
S(x+2\pi)=S(x).
\label{eq:Sx}
\end{equation}
The periodic energy density is then
\begin{equation}
\Ecal_{\rm cyl}(t,\theta)
\simeq \frac{1}{2\sqrt{2\pi}\,aL}
\left[p_u S(u)+p_v S(v)\right].
\label{eq:E-cyl}
\end{equation}

Equation \eqref{eq:E-cyl} describes two periodic Gaussian pulses
propagating in opposite directions along the boundary light-cone
directions.  To follow the physical pulses directly, consider the two
peaks that coincide at the insertion point
\((\tau,\theta)=(0,0)\), where \(\tau=t/L\).  Choosing the lifts that
initially satisfy \(u=v=0\), their centers obey
\begin{equation}
u=0
\quad\Longleftrightarrow\quad
\theta_u(\tau)=-\tau,
\qquad
v=0
\quad\Longleftrightarrow\quad
\theta_v(\tau)=+\tau,
\end{equation}
with the angular positions understood modulo \(2\pi\).  Thus the two
pulses propagate at the speed of light in opposite directions around
the spatial circle.  At \(\tau=\pi\), they reach
\begin{equation}
\theta_u=-\pi,
\qquad
\theta_v=+\pi.
\end{equation}
Because \(-\pi\) and \(+\pi\) denote the same point on the circle, the
two pulses meet at the antipodal point.  At \(\tau=2\pi\), each pulse
has completed one circuit of the cylinder and returns to the original
point.

The same propagation can be expressed in terms of the periodic images
on the covering space.  Their centers are located at
\begin{equation}
u=2\pi n_u,
\qquad
v=2\pi n_v,
\qquad
n_u,n_v\in\mathbb{Z}.
\label{eq:chiral-image-lattices}
\end{equation}
Here \(n_u\) and \(n_v\) label covering-space images of the same two
physical pulses; they do not label additional pulses on the cylinder.
A \(u\)-image and a \(v\)-image represent the same point on the
cylinder when
\begin{equation}
\tau=\pi(n_u+n_v),
\qquad
\theta=\pi(n_u-n_v)
\quad (\mathrm{mod}\ 2\pi).
\label{eq:chiral-intersections}
\end{equation}
For the first antipodal meeting, the representative \(\theta=\pi\)
is described by
\begin{equation}
(n_u,n_v)=(1,0),
\qquad
(\tau,\theta)=(\pi,\pi),
\end{equation}
whereas the equivalent representative \(\theta=-\pi\) is described by
\begin{equation}
(n_u,n_v)=(0,1),
\qquad
(\tau,\theta)=(\pi,-\pi).
\end{equation}
These are two covering-space descriptions of the same physical meeting
of the pulses that originated from \(u=v=0\).  The first return to the
original point is represented by
\begin{equation}
(n_u,n_v)=(1,1),
\qquad
(\tau,\theta)=(2\pi,0),
\end{equation}
for which \(u=v=2\pi\).  Thus the antipodal meeting and the return correspond to different
combinations of the same (u)- and (v)-chiral pole lattices.  The
different integer labels specify periodic images on the covering space
and do not represent additional physical pulses or an additional
family of poles.

The energy density remains spatially resolved between these events.
For example, at \(\tau=\pi/2\), while the corresponding bulk packet is
in the AdS interior, the two boundary peaks are centered at
\(\theta=-\pi/2\) and \(\theta=+\pi/2\).  This gives a continuously
evolving boundary observable, although the angular positions of the
peaks do not determine the radial position of the bulk packet.

The state-dependent information is instead contained in the residues,
or equivalently in the integrated chiral weights.  Define the
dimensionless charges over one spatial cycle by
\begin{equation}
Q_u:=L^2\int_{-\pi}^{\pi}\dd\theta\,\Ecal_u(t,\theta),
\qquad
Q_v:=L^2\int_{-\pi}^{\pi}\dd\theta\,\Ecal_v(t,\theta).
\label{eq:charges}
\end{equation}
Equation \eqref{eq:E-cyl} gives
\begin{equation}
Q_u=\frac{p_u}{2},
\qquad
Q_v=\frac{p_v}{2},
\qquad
Q_u+Q_v=\omega L,
\qquad
Q_v-Q_u=m_0.
\label{eq:chiral-charges}
\end{equation}
Consequently,
\begin{equation}
\frac{p_v-p_u}{p_v+p_u}
=\frac{m_0}{\omega L}
=b,
\label{eq:b-from-pu-pv}
\end{equation}
or equivalently
\begin{equation}
b=\frac{Q_v-Q_u}{Q_v+Q_u},
\qquad
\sin\rho_*=|b|.
\label{eq:b-from-chiral-charges}
\end{equation}
Null rays with different \(b\) have the same boundary endpoints and
arrival times, but different angular momenta and radial turning points.
\footnote{
    This degeneracy is special to pure AdS.  In a generic asymptotically
AdS geometry, gravitational time delay and angular deflection depend
on the bulk trajectory, so that the boundary arrival time and endpoint
can depend nontrivially on the impact parameter $b$; see, e.g.,
\cite{Gao:2000ga,Hubeny:2006yu}.
}
The image lattices fix the peak trajectories and their recurrence,
whereas the residues retain the state-dependent conserved data.  In
particular, \eqref{eq:b-from-pu-pv} identifies the relative chiral
weight with the impact parameter obtained independently from the bulk
analysis in Appendix \ref{app:bulk-wave-packet}.

\begin{figure}[t]
  \centering
  \includegraphics[width=.78\textwidth]{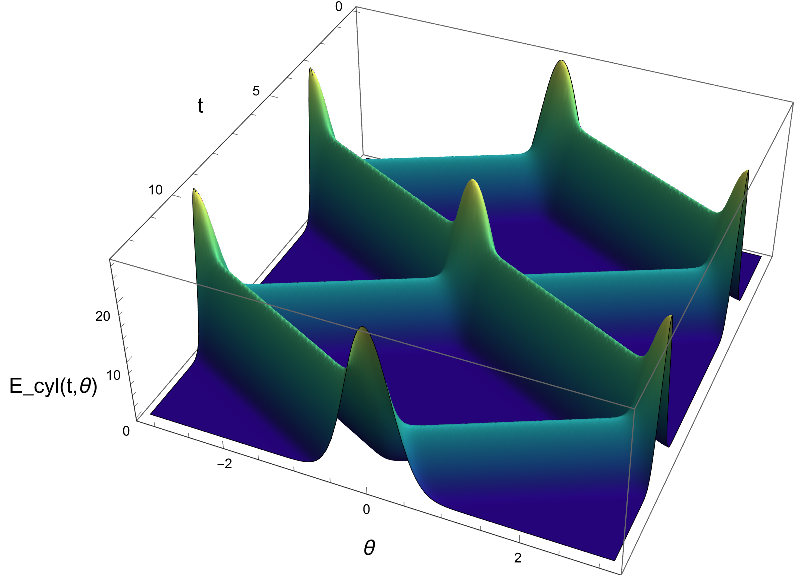}
    \caption{
    Three-dimensional plot of the periodic energy density \eqref{eq:E-cyl} for
    \(
    L=1,\ a=0.3,\ \omega=20,\ m_0=5
    \),
    with the periodic sum truncated at \(n_{\max}=10\), over the ranges
    \(
    0\le t\le 4\pi L
    \)
    and
    \(
    -\pi\le \theta\le \pi
    \).
    The two null peaks propagate along \(u=\mathrm{const.}\) and \(v=\mathrm{const.}\), and return periodically as required by the angular periodicity of the cylinder; this is consistent with the pole structure of the exact cylinder correlators.
    }
  \label{fig:wavepacket_3d}
\end{figure}

A typical example of the periodic energy density \eqref{eq:E-cyl} is shown in Fig.~\ref{fig:wavepacket_3d}.

\section{\texorpdfstring{LCD state and radial null propagation}{LCD state and radial null propagation}}
\label{sec:heavy-null}

We now give a second comparison between bulk propagation and boundary
energy density. In contrast to the wave-packet state of the previous
section, the state considered here is constructed from a local scalar primary
operator of large conformal dimension with a Euclidean time regulator. This is
the LCD state. The boundary stress tensor
can be computed exactly, and the resulting energy density consists of
two chiral pulses moving along the boundary light-cone directions. We
compare this boundary behavior with the bulk profile.

We consider the state
\begin{equation}
|\Psi_\epsilon\rangle
=
\mathcal{N}e^{-\epsilon H}\Ocal_\Delta(0)|0\rangle .
\label{eq:heavy-state}
\end{equation}
Here \(H\) denotes the dimensionless cylinder Hamiltonian generating
translations in the unit-cylinder time \(\tau\). Equivalently, if
\(H_{\rm phys}\) is the physical Hamiltonian, then
\(H=L H_{\rm phys}\), and the physical Euclidean time regulator is
\(L\epsilon\).
This construction is related to earlier studies of holographic local
quenches and local-operator excited states, where local CFT excitations
were analyzed together with their holographic descriptions
\cite{Nozaki:2013wia,Caputa:2014vaa}. The particular use of
primary insertions of large conformal dimension with Euclidean
time regulators as localized states in global AdS was developed in \cite{Berenstein:2019qmm}. More generally,
Euclidean sources provide a way to prepare Lorentzian semiclassical bulk
states in holographic CFTs \cite{Marolf:2017kvq}.

For large \(\Delta\), this state creates a localized
semiclassical bulk excitation in the probe regime. We take
\begin{equation}
1\ll\Delta\ll c,
\end{equation}
so that the dual bulk excitation admits a particle-like semiclassical
interpretation while its backreaction can be neglected.\footnote{
More precisely, in the UV-regulated state the typical dimensionless
energy is of order \(\omega_{\rm typ}\sim \Delta/\epsilon\). Thus the
probe approximation requires \(\omega_{\rm typ}\ll c\), or equivalently
\(\Delta/\epsilon\ll c\), rather than only \(\Delta\ll c\). We will
return to this point in Appendix~\ref{app:heavy-global-modes}.
}
The Euclidean time regulator \(\epsilon\) controls the UV energy of the
state. In the regime \(\epsilon\ll1\), the typical injected energy is
much larger than the rest energy of the dual bulk excitation, and the
leading bulk propagation is therefore governed by the null limit of
radial propagation. The key question is how this radial null behavior is
reflected in the boundary stress tensor.

For the CFT computation in this section, it is convenient to use the
unit-cylinder Euclidean coordinate
\begin{equation}
\zeta=\tau_E+ i\theta,
\qquad
\bar\zeta=\tau_E- i\theta .
\end{equation}
The expectation value of a local operator \(X(\zeta,\bar\zeta)\) is
\begin{equation}
\left\langle X(\zeta,\bar\zeta)\right\rangle_{\Psi_\epsilon}
=\frac{\left\langle
\Ocal^\dagger(\zeta_1,\bar\zeta_1)
X(\zeta,\bar\zeta)
\Ocal(\zeta_2,\bar\zeta_2)
\right\rangle_{\rm cyl}}
{\left\langle
\Ocal^\dagger(\zeta_1,\bar\zeta_1)
\Ocal(\zeta_2,\bar\zeta_2)
\right\rangle_{\rm cyl}},
\label{eq:heavy-expvalue}
\end{equation}
with
\begin{equation}
\zeta_1=+\epsilon,
\qquad
\zeta_2=-\epsilon .
\label{eq:zeta-labels}
\end{equation}
Here the ket insertion is at Euclidean time \(-\epsilon\), while the bra
insertion is at \(+\epsilon\).

\subsection{Exact cylinder stress tensor}

Using the plane-to-cylinder map $z=e^\zeta$, one has
\begin{equation}
z_1=e^{+\epsilon},
\qquad
z_2=e^{-\epsilon} .
\end{equation}
The plane Ward identity gives
\begin{equation}
\frac{\left\langle T(z)\Ocal(z_1,\bar z_1)\Ocal(z_2,\bar z_2)\right\rangle}
{\left\langle\Ocal(z_1,\bar z_1)\Ocal(z_2,\bar z_2)\right\rangle}
=h\frac{(z_1-z_2)^2}{(z-z_1)^2(z-z_2)^2} .
\end{equation}
The cylinder stress tensor is
\begin{equation}
T_{\rm cyl}(\zeta)=z^2T_{\rm plane}(z)-\frac{c}{24},
\end{equation}
and therefore
\begin{equation}
\left\langle T(\zeta)\right\rangle_{\Psi_\epsilon}
=\frac{h\sinh^2\epsilon}{4\sinh^2\frac{\zeta+\epsilon}{2}\sinh^2\frac{\zeta-\epsilon}{2}}
-\frac{c}{24} .
\label{eq:T-heavy-eucl}
\end{equation}
Similarly,
\begin{equation}
\left\langle \bar T(\bar\zeta)\right\rangle_{\Psi_\epsilon}
=\frac{h\sinh^2\epsilon}{4\sinh^2\frac{\bar\zeta+\epsilon}{2}\sinh^2\frac{\bar\zeta-\epsilon}{2}}
-\frac{c}{24} .
\end{equation}
Subtracting the cylinder vacuum contribution,
\begin{equation}
\langle T\rangle_0=\langle\bar T\rangle_0=-\frac{c}{24},
\end{equation}
we obtain
\begin{equation}
\delta\langle T(\zeta)\rangle
=\frac{h\sinh^2\epsilon}{4\sinh^2\frac{\zeta+\epsilon}{2}\sinh^2\frac{\zeta-\epsilon}{2}},
\qquad
\delta\langle \bar T(\bar\zeta)\rangle
=\frac{h\sinh^2\epsilon}{4\sinh^2\frac{\bar\zeta+\epsilon}{2}\sinh^2\frac{\bar\zeta-\epsilon}{2}} .
\end{equation}

\subsection{Lorentzian energy density and comparison with the bulk profile}

After Lorentzian continuation
\begin{equation}
\zeta\to  i(\tau+\theta),
\qquad
\bar\zeta\to  i(\tau-\theta),
\end{equation}
one finds
\begin{equation}
\delta\langle T(\tau+\theta)\rangle
=\frac{h\sinh^2\epsilon}{4\left(\sin^2\frac{\tau+\theta}{2}+\sinh^2\frac{\epsilon}{2}\right)^2},
\qquad
\delta\langle\bar T(\tau-\theta)\rangle
=\frac{h\sinh^2\epsilon}{4\left(\sin^2\frac{\tau-\theta}{2}+\sinh^2\frac{\epsilon}{2}\right)^2} .
\end{equation}
Hence the vacuum-subtracted stress-tensor component on the unit cylinder is
\begin{equation}
\delta\langle T_{\tau\tau}(\tau,\theta)\rangle
=\frac{\Delta\sinh^2\epsilon}{8}
\left[
\frac{1}{\left(\sin^2\frac{\tau+\theta}{2}+\sinh^2\frac{\epsilon}{2}\right)^2}
+
\frac{1}{\left(\sin^2\frac{\tau-\theta}{2}+\sinh^2\frac{\epsilon}{2}\right)^2}
\right] .
\label{eq:Ttautau-heavy}
\end{equation}
The corresponding energy density is
\begin{equation}
\Ecal_\epsilon^{(\tau)}(\tau,\theta)
:=\frac{1}{2\pi}\delta\langle T_{\tau\tau}(\tau,\theta)\rangle .
\end{equation}
This exact cylinder result exhibits two chiral pulses moving continuously along the boundary light-cone directions
\begin{equation}
\tau+\theta=2\pi n,
\qquad
\tau-\theta=2\pi n,
\qquad
n\in\mathbb Z .
\end{equation}
The two pulses meet at the antipodal point after
\begin{equation}
\Delta\tau=\pi
\end{equation}
and return to the original point after one full global period,
\begin{equation}
\Delta\tau=2\pi .
\end{equation}
For every finite \(\epsilon>0\), equation
\eqref{eq:Ttautau-heavy} is a smooth, nonvanishing boundary
distribution throughout the evolution.  The two chiral pulses can
therefore be followed continuously while the bulk excitation
propagates through the AdS interior.  They meet at the antipodal point
at \(\tau=\pi\) and at the original point at \(\tau=2\pi\), in agreement
with the boundary-arrival times of radial null propagation in global
\(\AdS_3\).

The dimensionless CFT energy is obtained from the cylinder Hamiltonian as
\begin{equation}
\Omega_{\CFT}
=\int_{-\pi}^{\pi}\dd\theta\,\Ecal_\epsilon^{(\tau)}(\tau,\theta)
=\frac{1}{2\pi}\int_{-\pi}^{\pi}\dd\theta\,\delta\langle T_{\tau\tau}(\tau,\theta)\rangle
=\Delta\coth\epsilon .
\label{eq:LCD-energy}
\end{equation}
This exact result gives a useful check on the energy scale of the state.  The bulk interpretation, however, is most directly seen from the bulk
wave function
\begin{equation}
\Psi_\epsilon(\tau,\rho,\Omega;\Omega_0)
=
\langle0|\Phi(\tau,\rho,\Omega)e^{-\epsilon H}
\Ocal_\Delta(0,\Omega_0)|0\rangle .
\end{equation}
As shown in Appendix \ref{app:heavy-global-modes}, in the large-$\Delta$ and small-$\epsilon$ regime, the regulated bulk profile describes a localized excitation propagating along the radial null geodesic. For the incoming branch, $\Omega=\Omega_0$, the trajectory is
\begin{equation}
\tau=\frac{\pi}{2}-\rho . 
\end{equation}
The same global-mode sum shows that the typical dimensionless energy in the UV-regulated regime is
\begin{equation}
\omega_{\rm typ}\sim \frac{\Delta}{\epsilon} .
\end{equation}
Restoring the AdS radius, this corresponds to
\begin{equation}
E_{\rm typ}^{\rm phys}\sim \frac{\Delta}{L\epsilon} .
\end{equation}
Since the dimensionless rest energy of the dual bulk excitation is
\(ML\simeq\Delta\), we have
\begin{equation}
\frac{E_{\rm typ}^{\rm phys}}{M}
=
\frac{\omega_{\rm typ}}{ML}
\sim
\frac{1}{\epsilon}
\gg1
\end{equation}
for \(\epsilon\ll1\).
Together with the direct peak-locus calculation in Appendix
\ref{app:heavy-global-modes}, this shows that the leading bulk
propagation approaches the radial null geodesic.  This is the \(m_0=0\) case of the null
wave-packet trajectory discussed in Section \ref{sec:null-wave}. 
The exact boundary energy density exhibits
the same sequence of boundary arrivals through its two chiral pulses.

\section{Discussion}
\label{sec:discussion}

\paragraph{Boundary signatures of null propagation.}
The two examples considered in this paper have different detailed
structures, but they convey a common message: the global-cylinder
description makes visible a boundary-to-boundary structure that is not
manifest in the Poincar\'e-patch analysis. In global \(\AdS_3\), null rays
emitted from the boundary reach the antipodal boundary point and return
to the original point after one global period. We have shown that this
bulk periodicity is reflected in boundary energy densities as periodic
chiral structures on the cylinder.
The \(2\pi\) periodicity itself follows from the angular identification
and the image structure of the cylinder correlators.  The comparison
with the bulk propagation gives a more detailed result: the two chiral
peaks propagate continuously and the boundary energy density remains
nonzero while the bulk wave packet is in the AdS interior.  The peaks
meet at the antipodal point after \(\Delta\tau=\pi\), and their
state-dependent weights reproduce the conserved data of the bulk null
ray.
The boundary energy density contains no bulk radial coordinate and
does not directly reconstruct the local radial position of the wave packet.
Independently, the radial trajectory is computed on the bulk side in the
appendices.  The CFT calculation instead gives the time-dependent
boundary energy distribution, including its behavior at intermediate
times, its antipodal and return times, and its integrated energy and
angular momentum.

For the wave-packet state, the local residue analysis yields two
Gaussian peaks localized near \(u=0\) and \(v=0\). In the Poincar\'e-patch
analysis these peaks describe the local light-cone propagation of the
boundary energy density. On the global cylinder, the exact correlators
contain the corresponding lattice of light-cone poles, and the leading
small-width result is obtained by summing the local pole contributions
near these poles. Thus the two local peaks are promoted to periodic
chiral peaks on the cylinder. Their propagation reproduces the arrival
at the antipodal boundary point and the return to the original boundary
point.  The result also makes clear
how the bulk impact parameter is encoded in the CFT observable. The
locations of the null peaks in the dual CFT are independent of the
impact parameter, while the relative weights of the two chiral peaks
encode
\[
\frac{p_v-p_u}{p_v+p_u}=\frac{m_0}{\omega L}=b .
\]
More precisely, the integrated weights satisfy
\[
Q_u+Q_v=\omega L,
\qquad
Q_v-Q_u=m_0,
\qquad
b=\frac{Q_v-Q_u}{Q_v+Q_u}.
\]
Thus wave packets with different energy and angular momentum have the
same pole locations but different weights.  The boundary energy density
therefore retains the information that distinguishes null rays with
different impact parameters and radial turning points.
In this sense, the boundary energy density captures part of the
conserved data of the bulk null ray, even though it does not literally
draw the bulk trajectory.

For the LCD state, the exact
cylinder computation gives two periodic pulses moving along the
boundary light-cone directions. This
provides a second realization of the same global feature: the boundary
pulses reach the antipodal point and return to the original point with
the same global periodicity as the radial null geodesic described by the
regulated bulk profile. The total dimensionless energy is
\[
\Omega_{\CFT}=\Delta\coth\epsilon .
\]
Since the rest energy of the dual bulk excitation is \(ML\simeq\Delta\),
the relevant ratio for the local insertion with
\(\epsilon\ll1\) is
\begin{equation}
\frac{E_{\rm phys}}{M}
\simeq
\frac{1}{\epsilon}
\gg1 .
\end{equation}
The global-mode calculation in Appendix
\ref{app:heavy-global-modes} further shows that the peak of the bulk
wave function approaches the radial null trajectory.  The
exact boundary energy density remains nonzero throughout the
evolution, and its two pulses meet at the antipodal point after
\(\Delta\tau=\pi\) and return to the original point after
\(\Delta\tau=2\pi\).  Thus the small-\(\epsilon\) limit reproduces
radial null propagation.
This agrees with the \(m_0=0\) limit of the wave-packet construction.

Thus, in both examples, the leading bulk propagation relevant for the
boundary energy density is described by null geodesics in global AdS.
The wave-packet state realizes a general null geodesic with impact
parameter \(b=m_0/(\omega L)\), whereas the LCD state corresponds
to the radial case \(b=0\). The common \(2\pi\) periodicity and the
locations of the repeated light-cone poles are fixed by the cylinder
kinematics.  The agreement with the bulk description includes the
meeting of the two chiral peaks at \(\Delta\tau=\pi\), the nonzero
energy-density profile at intermediate times, and, for the wave-packet
state, the energy and angular-momentum dependence of the two weights.

\paragraph{Relation to energy-flow and light-ray observables.}
The local stress tensor studied here is also the basic ingredient of
energy-flow observables and stress-tensor light-ray operators, including
the averaged null-energy operator
\cite{Hofman:2008ar,Kravchuk:2018htv,Faulkner:2016mzt}.  These
observables extract integrated information from the stress tensor,
whereas our calculation retains the local cylinder profile
\begin{equation}
T_{tt}(t,\theta)=T(u)+\overline{T}(v),
\end{equation}
and therefore resolves the positions and time evolution of the two
chiral pulses.  The one-cycle integrals $Q_u$ and $Q_v$ defined in
\eqref{eq:charges} measure the total left- and right-moving energies.
Their sum and difference give the total energy and angular momentum,
and their normalized difference gives the bulk impact parameter.

The one-cycle integrals considered here are distinct from the standard
averaged null-energy operator, which is defined by integrating a null
component of the stress tensor along a complete null line.  The global
structure is also different on the cylinder, where the chiral pulses
recur periodically rather than propagating along a noncompact null
direction.  The local profile and the one-cycle chiral integrals
therefore retain complementary information: the former resolves the
temporal and angular evolution of the pulses, while the latter retain
the conserved energy and angular momentum of the corresponding bulk
excitation.

\paragraph{Relation to Lorentzian singularities.}
Our results are also closely related to the broader idea that bulk
causal relations leave characteristic signatures in Lorentzian boundary
observables. A basic example is provided by bulk-cone singularities,
where Lorentzian CFT correlators become singular when boundary points
are connected by bulk null geodesics \cite{Hubeny:2006yu}. Related
bulk-cone singularities have also been studied in black-hole and
horizonless AdS geometries
\cite{Dodelson:2023nnr,Chen:2025jbf}. This is also connected to
bulk-point singularities and bulk locality in CFT correlators
\cite{Bousso:2012mh,Maldacena:2015iua,Kinoshita:2023lgy}.

\paragraph{Extensions to nontrivial bulk geometries.}
A particularly interesting extension is to introduce a heavy particle
in the bulk, corresponding to a conical-defect background.  In contrast
to pure global AdS, null propagation in conical AdS exhibits a
nontrivial time delay.  It is therefore natural to ask how this delayed
bulk propagation is reflected in boundary observables.  Since the
corresponding CFT calculation involves a heavy-light state and is no
longer determined solely by the vacuum-cylinder kinematics, it may
provide a genuinely nontrivial test of the sensitivity of boundary
energy densities to the bulk geometry. The conical-defect geometry is a natural horizonless extension of the
present setup, because null rays can still return to the boundary.
A qualitatively different question is how the global-AdS periodic
structure is modified when a horizon is present.

Wave packets in a planar BTZ black hole have already been studied in
\cite{Tanahashi:2026mia}, where it was found that, because of special
features of the BTZ geometry and two-dimensional conformal symmetry,
the gravitational time delay is not directly visible in the boundary
energy density.  This suggests that, even in global BTZ, the energy
density alone may not be a sufficiently sensitive observable for
extracting bulk time delay.  Moreover, unlike in global AdS, the global
BTZ geometry does not support the same periodic boundary-to-boundary
propagation of null rays: the effective potential for null geodesics
has no maximum that would turn the ray back toward the boundary, so
inward-directed null rays entering from the boundary fall into the
black hole \cite{Kinoshita:2023lgy}.
A global-BTZ analysis would therefore address how the pure-AdS pole
structure and periodic energy density are modified or lost in the
presence of a horizon, rather than providing a direct extension of the
periodic propagation studied here.

Both the conical-defect and black-hole extensions also raise the
question of gravitational backreaction.
The present analysis is restricted to the probe regime.  Beyond this
regime, gravitational backreaction must be included.  A natural
approach in the bulk is then to solve the time-dependent bulk Einstein
equations and extract the boundary stress tensor from the near-boundary
metric using holographic renormalization
\cite{Balasubramanian:1999re,deHaro:2000vlm}, as in numerical studies
of asymptotically AdS collapse and black-hole formation
\cite{Bantilan:2012vu}.  This would provide a way to study boundary
energy densities when the injected energy is large enough to modify the
bulk geometry.
Another complementary direction is to construct
the corresponding high-energy states directly in the CFT. In this
respect, it is interesting that bulk wave packets have recently been
used to construct CFT states describing black-hole formation and
evaporation in AdS/CFT \cite{deBoer:2026cng}. It would be useful to
understand how boundary energy densities behave in such wave-packet states with gravitational backreaction, where the injected energy is large enough to modify
the bulk geometry.

\paragraph{Higher-dimensional generalizations.}
Another direction is to understand which aspects of the present result
are special to \(\AdS_3/\CFT_2\). The chiral peak structure found
here relies strongly on two-dimensional conformal symmetry and the
decomposition into left- and right-moving sectors. In higher-dimensional
examples, the boundary energy density associated with a bulk wave packet
can be more spatially spread \cite{Tanahashi:2025fqi}. Comparing the
present global-cylinder analysis with higher-dimensional wave packets
would help separate universal signatures of bulk propagation from
features special to two-dimensional CFTs.

\section*{Acknowledgments}

The author would like to thank R.~Adachi, K.~Hashimoto, T.~Kawamoto,
N.~Tanahashi, K.~Tasuki and S.~Terashima for valuable discussions and helpful
comments. The author is particularly grateful to S.~Terashima for
detailed discussions on the bulk interpretation of the LCD state, which were essential for Appendix \ref{app:heavy-global-modes}.

During the preparation of this manuscript, the author used ChatGPT
(OpenAI) to assist with language editing, organization of the text, and
checking the clarity of explanations. The author reviewed and edited
all AI-assisted output and takes full responsibility for the content of
the manuscript.

\appendix

\section{Bulk wave packet from global Klein--Gordon modes}
\label{app:bulk-wave-packet}

In this appendix, we check the bulk interpretation of the boundary-smeared state used in the main text. We show that the state prepares a bulk one-particle wave packet in the global patch and that, in the high-frequency and small-width regime, its bulk profile is localized around a null geodesic in global $\AdS_3$.

\subsection{Global modes and the wave-packet state}

We consider a scalar field $\Phi$ of mass $M$ in global $\AdS_3$,
\begin{equation}
\dd s^2=\frac{L^2}{\cos^2\rho}\left(-\dd\tau^2+\dd\rho^2+\sin^2\rho\,\dd\theta^2\right),
\qquad
0\leq\rho<\frac{\pi}{2},
\qquad
\theta\sim\theta+2\pi .
\end{equation}
Here $\tau$ is the dimensionless global time, and the corresponding dimensionful boundary time is $t=L\tau$.

The normalizable global modes of a scalar field are standard; see, for
example, \cite{Balasubramanian:1998sn}. We write them as
\begin{equation}
u_{n,m}(\tau,\rho,\theta)=\mathcal{N}_{n,m}e^{- i\Omega_{n,m}\tau+ i m\theta}R_{n,m}(\rho),
\end{equation}
where
\begin{equation}
R_{n,m}(\rho)=(\cos\rho)^\Delta(\sin\rho)^{|m|}P_n^{(|m|,\Delta-1)}(\cos2\rho),
\end{equation}
and
\begin{equation}
\Omega_{n,m}=\Delta+|m|+2n,
\qquad
n=0,1,2,\ldots,
\qquad
m\in\mathbb{Z} .
\end{equation}
Here $P_n^{(\alpha,\beta)}$ denotes a Jacobi polynomial, and we use the standard quantization
\begin{equation}
\Delta=1+\sqrt{1+M^2L^2} .
\end{equation}
The quantized bulk field is expanded as
\begin{equation}
\Phi(\tau,\rho,\theta)
=\sum_{n=0}^{\infty}\sum_{m\in\mathbb{Z}}
\left(a_{n,m}u_{n,m}(\tau,\rho,\theta)+a_{n,m}^\dagger u^*_{n,m}(\tau,\rho,\theta)\right) .
\end{equation}

We now consider the boundary wave-packet state used in the main text,
\begin{equation}
|\Omega_0,m_0\rangle
=\int_{-\infty}^{\infty}\dd\tau\int_{-\pi}^{\pi}\dd\theta\,
 f_{\Omega_0,m_0}(\tau,\theta)\Ocal(\tau,\theta)|0\rangle,
\end{equation}
with
\begin{equation}
f_{\Omega_0,m_0}(\tau,\theta)
=\exp\left[-\frac{\tau^2+4\tan^2(\theta/2)}{2\alpha^2}- i\Omega_0\tau+ i m_0\theta\right],
\qquad
\alpha=\frac{a}{L} .
\end{equation}
Here $\Omega_0=\omega L$ is the dimensionless central frequency and $m_0$ is the central angular momentum. The extrapolate dictionary then relates the boundary primary to the
near-boundary limit of the bulk field
\cite{Balasubramanian:1998sn,Banks:1998dd,Hamilton:2006az},
\begin{equation}
\Ocal(\tau,\theta)\sim \lim_{\rho\to\pi/2}(\cos\rho)^{-\Delta}
\Phi(\tau,\rho,\theta) .
\end{equation}
Therefore the action of the boundary operator on the vacuum can be written as
\begin{equation}
\Ocal(\tau,\theta)|0\rangle
=\sum_{n,m}B_{n,m}e^{ i\Omega_{n,m}\tau- i m\theta}a^\dagger_{n,m}|0\rangle .
\label{eq:boundary-mode-expansion}
\end{equation}
We choose the phase convention of the modes so that $B_{n,m}$ is a slowly varying envelope in the large quantum-number regime. Any residual rapidly varying phase in $B_{n,m}$ can equivalently be absorbed into the eikonal action used below, and does not affect the stationary-phase trajectory.

Substituting \eqref{eq:boundary-mode-expansion} into the smeared state, we obtain a bulk one-particle state
\begin{equation}
|\Omega_0,m_0\rangle
=\sum_{n,m}c_{n,m}a^\dagger_{n,m}|0\rangle,
\end{equation}
where
\begin{equation}
c_{n,m}\propto B_{n,m}\int\dd\tau\dd\theta\,
 f_{\Omega_0,m_0}(\tau,\theta)e^{ i\Omega_{n,m}\tau- i m\theta} .
\end{equation}
For $\alpha\ll1$, the smearing is localized near $\theta=0$, so that $2\tan(\theta/2)\simeq\theta$, and the angular integral over the fundamental domain can be extended to the real line in the Gaussian approximation. We then find
\begin{equation}
c_{n,m}\propto B_{n,m}
\exp\left[-\frac{\alpha^2}{2}(\Omega_{n,m}-\Omega_0)^2
-\frac{\alpha^2}{2}(m-m_0)^2\right] .
\end{equation}
Thus the CFT state prepared by the smearing is, in the bulk one-particle Hilbert space, a superposition of global modes sharply peaked around
\begin{equation}
\Omega_{n,m}\simeq\Omega_0,
\qquad
m\simeq m_0 .
\end{equation}

\subsection{Bulk profile and stationary phase}

To see how this state propagates in the bulk, we consider the
time-dependent bulk profile
\begin{equation}
\Psi_{\rm wp}(T,\rho,\theta)
\equiv
\langle0|\Phi(T,\rho,\theta)|\Omega_0,m_0\rangle
=\langle0|\Phi(0,\rho,\theta)e^{- i HT}|\Omega_0,m_0\rangle,
\label{eq:bulk-profile}
\end{equation}
where \(T\) is the elapsed dimensionless global time.\footnote{This
quantity is the analogue of a one-particle wave function in QFT: it is
obtained by taking the overlap of the state with the bulk field
operator at a spacetime point. See, for example,
\cite{Shibuya:2025rel} for a related use of such a wave function.}
Using the mode expansion,
\begin{equation}
\Psi_{\rm wp}(T,\rho,\theta)
=\sum_{n,m}c_{n,m}\mathcal{N}_{n,m}R_{n,m}(\rho)e^{- i\Omega_{n,m}T+ i m\theta} .
\end{equation}
We evaluate this expression in the high-frequency wave-packet regime
\begin{equation}
\alpha\ll1,
\qquad
\Omega_0\gg\Delta,
\qquad
\alpha(\Omega_0-m_0)\gg1,
\qquad
\alpha(\Omega_0+m_0)\gg1,
\qquad
\Omega_0>|m_0| .
\label{eq:app-regime}
\end{equation}
The condition \(\alpha\ll1\) means that the wave packet is localized on
a scale much smaller than the AdS radius. The inequalities
\(\alpha(\Omega_0-m_0)\gg1\) and
\(\alpha(\Omega_0+m_0)\gg1\) ensure that the smearing contains many
oscillations in both light-cone directions. The resulting bulk wave
packet therefore has a well-defined central energy \(\Omega_0\) and
angular momentum \(m_0\), with the direction of angular motion fixed by
the sign of \(m_0\). The last condition, \(\Omega_0>|m_0|\), ensures
that the central ray has a classically allowed radial region, with a
turning point satisfying \(\sin\rho_*=|m_0|/\Omega_0\).

We now determine the leading phase of the radial wave function. This phase is fixed by the leading eikonal limit of the Klein--Gordon equation. Writing the rapidly varying part of a mode as
\begin{equation}
\exp\left[i I(T,\rho,\theta)\right],
\qquad
I(T,\rho,\theta)
=
-\Omega T+m(\theta-\theta_0)+W(\rho),
\end{equation}
where \(\theta_0\) denotes the angular position of the emitted ray. For
the state centered at \(\theta=0\) in the main text, one should set
\(\theta_0=0\). The leading Hamilton--Jacobi equation in global
\(\AdS_3\) is
\begin{equation}
-\Omega^2+\left(\partial_\rho W\right)^2
+\frac{m^2}{\sin^2\rho}
+\frac{M^2L^2}{\cos^2\rho}=0 .
\label{eq:HJ-massive}
\end{equation}
In the regime \(\Omega_0\gg\Delta\), the mass term gives subleading corrections to the central high-frequency ray in the bulk interior. It remains important for the near-boundary falloff of the exact normalizable mode, but it does not affect the leading null trajectory of the wave packet. Thus, for the purpose of determining the central ray, we use the leading null eikonal equation
\begin{equation}
-\Omega^2+\left(\partial_\rho W\right)^2
+\frac{m^2}{\sin^2\rho}=0 .
\label{eq:HJ-null}
\end{equation}
This gives
\begin{equation}
\partial_\rho W=\pm p_\rho(\rho;\Omega,m),
\qquad
p_\rho(\rho;\Omega,m)
=
\sqrt{\Omega^2-\frac{m^2}{\sin^2\rho}} .
\label{eq:prho-eikonal}
\end{equation}

The sign in \eqref{eq:prho-eikonal} labels the two WKB branches. The branch describing a future-directed ray entering the bulk from the boundary at \(T=0\) has negative radial momentum. We therefore choose the incoming Hamilton--Jacobi action
\begin{equation}
S_{\rm in}(\rho;\Omega,m)
=
\int_\rho^{\pi/2}\dd\rho'\,
\sqrt{\Omega^2-\frac{m^2}{\sin^2\rho'}} .
\label{eq:S-in}
\end{equation}
Indeed,
\begin{equation}
\partial_\rho S_{\rm in}(\rho;\Omega,m)
=
-\sqrt{\Omega^2-\frac{m^2}{\sin^2\rho}},
\end{equation}
so this branch has negative radial momentum and hence moves from the boundary toward smaller \(\rho\). In the WKB approximation the corresponding branch of the radial wave function is
\begin{equation}
R_{n,m}^{\rm in}(\rho)
\sim A_{\rm WKB}(\rho;\Omega,m)
\exp\left[iS_{\rm in}(\rho;\Omega,m)\right].
\end{equation}
We combine the WKB amplitude, the mode normalization, and the slowly varying part of the smearing coefficient into a single smooth prefactor \(B_{\rm in}(\rho;\Omega,m)\). This prefactor affects the overall amplitude of the wave packet at leading order, while the center of the packet is determined by the rapidly varying phase.

Keeping this incoming WKB branch, the bulk profile is locally
\begin{align}
\Psi^{\rm in}_{\rm wp}
&\sim \int \dd\Omega\,\dd m\,
\exp\left[-\frac{\alpha^2}{2}(\Omega-\Omega_0)^2
-\frac{\alpha^2}{2}(m-m_0)^2\right]
B_{\rm in}(\rho;\Omega,m)
\notag\\
&\quad\times
\exp\left[
-i\Omega T+i m(\theta-\theta_0)
+iS_{\rm in}(\rho;\Omega,m)
\right] .
\label{eq:wkb-integral}
\end{align}
Here the discrete sum over \((n,m)\) has been approximated by a local integral over \((\Omega,m)\). This is justified in the large quantum-number regime because the wave packet contains many modes within its frequency and angular-momentum widths.

Let
\begin{equation}
\delta\Omega=\Omega-\Omega_0,
\qquad
\delta m=m-m_0 .
\end{equation}
Expanding the phase around \((\Omega_0,m_0)\),
\begin{equation}
S_{\rm in}(\rho;\Omega,m)
=
S_{{\rm in},0}
+\delta\Omega\,\partial_\Omega S_{{\rm in},0}
+\delta m\,\partial_m S_{{\rm in},0}
+\cdots ,
\end{equation}
where
\begin{equation}
S_{{\rm in},0}:=S_{\rm in}(\rho;\Omega_0,m_0),
\qquad
B_{{\rm in},0}:=B_{\rm in}(\rho;\Omega_0,m_0),
\end{equation}
we obtain
\begin{align}
\Psi^{\rm in}_{\rm wp}
&\sim B_{{\rm in},0}e^{ i\Theta_0}
\int \dd\delta\Omega\dd\delta m\,
\exp\left[
-\frac{\alpha^2}{2}(\delta\Omega^2+\delta m^2)
+i\delta\Omega\left(-T+\partial_\Omega S_{{\rm in},0}\right)
\right.
\notag\\
&\hspace{8em}\left.
+i\delta m\left(\theta-\theta_0+\partial_m S_{{\rm in},0}\right)
\right],
\end{align}
with
\begin{equation}
\Theta_0
=
-\Omega_0T+m_0(\theta-\theta_0)+S_{{\rm in},0}.
\end{equation}
Performing the Gaussian integral gives
\begin{equation}
\Psi^{\rm in}_{\rm wp}
\sim B_{{\rm in},0}e^{ i\Theta_0}
\exp\left[
-\frac{
\left(T-\partial_\Omega S_{{\rm in},0}\right)^2
+
\left(\theta-\theta_0+\partial_m S_{{\rm in},0}\right)^2
}{2\alpha^2}
\right] .
\end{equation}
Thus this incoming branch is exponentially supported on
\begin{equation}
T=\partial_\Omega S_{{\rm in},0},
\qquad
\theta-\theta_0=-\partial_m S_{{\rm in},0}.
\label{eq:stationary-phase}
\end{equation}
These are precisely the Hamilton--Jacobi relations for the null ray generated by \(S_{\rm in}\). 

\subsection{Null geodesic from the stationary-phase conditions}

The stationary-phase conditions above are precisely the Hamilton--Jacobi relations for the incoming branch of a null ray. Defining
\begin{equation}
b=\frac{m_0}{\Omega_0},
\end{equation}
we have
\begin{equation}
p_\rho(\rho;\Omega_0,m_0)
=\Omega_0\sqrt{1-\frac{b^2}{\sin^2\rho}} .
\end{equation}
The derivatives of the incoming Hamilton--Jacobi action give
\begin{equation}
\partial_\Omega S_{{\rm in},0}
=
\int_\rho^{\pi/2}\dd\rho'\,
\frac{1}{\sqrt{1-\frac{b^2}{\sin^2\rho'}}},
\label{eq:dOmegaS}
\end{equation}
and
\begin{equation}
\partial_m S_{{\rm in},0}
=
-\int_\rho^{\pi/2}\dd\rho'\,
\frac{b}{\sin^2\rho'\sqrt{1-\frac{b^2}{\sin^2\rho'}}} .
\label{eq:dmS}
\end{equation}
These derivatives have the standard Hamilton--Jacobi interpretation.
The condition
\begin{equation}
T=\partial_\Omega S_{{\rm in},0}
\end{equation}
gives the elapsed global time from the boundary to the radial position
\(\rho\), while
\begin{equation}
\theta-\theta_0=-\partial_m S_{{\rm in},0}
\end{equation}
gives the angular displacement of the center of the wave packet. The
minus sign in the second relation follows from our phase convention
\(\exp[i m(\theta-\theta_0)]\) in \eqref{eq:wkb-integral}.

Let us also derive the corresponding differential form of the trajectory.
The incoming Hamilton--Jacobi action is
\begin{equation}
S_{\rm in}(\rho;\Omega,m)
=
\int_\rho^{\pi/2}\dd\rho'\,
\sqrt{\Omega^2-\frac{m^2}{\sin^2\rho'}} .
\end{equation}
Therefore the radial momentum of the incoming branch is
\begin{equation}
p_\rho
=
\partial_\rho S_{\rm in}
=
-\sqrt{\Omega^2-\frac{m^2}{\sin^2\rho}} .
\label{eq:prho-incoming}
\end{equation}
The negative sign is the statement that the ray is moving inward, from
the boundary toward smaller \(\rho\). The leading eikonal equation is
the null Hamilton--Jacobi equation
\begin{equation}
-\Omega^2+p_\rho^2+\frac{m^2}{\sin^2\rho}=0 .
\label{eq:null-HJ-app}
\end{equation}
This is the Hamiltonian constraint for a null geodesic in the conformal
metric
\[
-\dd T^2+\dd\rho^2+\sin^2\rho\,\dd\theta^2 .
\]
The conserved quantities are
\begin{equation}
\Omega=-p_T,
\qquad
m=p_\theta .
\end{equation}
Using the global time \(T\) as the parameter along the ray, the
Hamilton equations give
\begin{equation}
\frac{\dd\rho}{\dd T}
=
\frac{p_\rho}{\Omega},
\qquad
\frac{\dd\theta}{\dd T}
=
\frac{m}{\Omega\sin^2\rho}.
\label{eq:geo-from-HJ}
\end{equation}
Defining
\begin{equation}
b=\frac{m}{\Omega},
\end{equation}
we therefore obtain
\begin{equation}
\left(\frac{\dd\rho}{\dd T}\right)^2
=
1-\frac{b^2}{\sin^2\rho},
\qquad
\frac{\dd\theta}{\dd T}
=
\frac{b}{\sin^2\rho}.
\label{eq:geo-diff-app}
\end{equation}
For the incoming branch one takes the negative sign of
\(\dd\rho/\dd T\). The turning point is determined by
\begin{equation}
\frac{\dd\rho}{\dd T}=0,
\end{equation}
and hence
\begin{equation}
\sin\rho_*=|b| .
\end{equation}
Thus the ratio of the central angular momentum and energy of the packet
fixes the radial depth of the null trajectory in the bulk.

For a wave packet emitted from the boundary at \(T=0\) and \(\theta=\theta_0\), a convenient parametrization of the corresponding null geodesic is
\begin{equation}
\cos\rho(T)=\sqrt{1-b^2}\,\sin T,
\qquad
0\leq T\leq\pi,
\end{equation}
and, for \(b>0\),
\begin{equation}
\theta(T)=\theta_0+
\begin{cases}
\arctan(b\tan T), & 0\leq T\leq\pi/2,\\[0.4em]
\pi+\arctan(b\tan T), & \pi/2\leq T\leq\pi .
\end{cases}
\end{equation}
For \(b<0\), the angular displacement has the opposite orientation, but the endpoint is the same antipodal point modulo \(2\pi\). Hence the packet reaches the turning point at \(T=\pi/2\), arrives at the antipodal boundary point at \(T=\pi\), and returns to the original boundary point after one global period,
\(T=2\pi\). In terms of the dimensionful boundary time \(t=LT\),
\begin{equation}
\Delta t^{\rm null}_{\rm bulk}=\pi L,
\qquad
T_{\rm period}=2\pi,
\qquad
t_{\rm period}=2\pi L .
\end{equation}
Finally, in terms of the boundary parameters used in the main text,
\begin{equation}
\Omega_0=\omega L,
\qquad
b=\frac{m_0}{\Omega_0}=\frac{m_0}{\omega L} .
\end{equation}
Thus the regime
\begin{equation}
a\ll L,
\qquad
\frac{(\omega L-m_0)a}{L}\gg1,
\qquad
\frac{(\omega L+m_0)a}{L}\gg1,
\qquad
\omega L>|m_0|
\end{equation}
is the regime in which the boundary-smeared state creates a localized bulk wave packet following a null geodesic in global \(\AdS_3\).

\section{\texorpdfstring{Bulk interpretation of the LCD state}{Bulk interpretation of the LCD state}}
\label{app:heavy-global-modes}

In this appendix we examine the bulk interpretation of the state
constructed from a local scalar primary insertion of large conformal
dimension with a Euclidean time regulator. We refer to it as the LCD state.
We work in the probe regime. The operator dimension is taken to satisfy
\(1\ll\Delta\), so that the dual bulk excitation admits a semiclassical
particle interpretation, but its backreaction is neglected. 
Since the global frequencies used below are dimensionless, we use the
dimensionless global time \(\tau\) in this appendix. The Hamiltonian
\(H\) in \(e^{-\epsilon H}\) is the corresponding dimensionless
cylinder Hamiltonian, and \(\epsilon\) is a dimensionless Euclidean time
regulator. Equivalently, if \(H_{\rm phys}\) is the physical
Hamiltonian, then \(H=L H_{\rm phys}\), and the physical Euclidean time
regulator is \(L\epsilon\).

Using global AdS normal modes, we compute the regulated bulk profile
\begin{equation}
\Psi_\epsilon(\tau,\rho,\hat n;\hat n_0)
:=
\langle0|\Phi(\tau,\rho,\hat n)e^{-\epsilon H}
\Ocal_\Delta(0,\hat n_0)|0\rangle .
\label{eq:app-heavy-profile-def}
\end{equation}
This quantity is an unnormalized bulk wave function associated with
the LCD state. The overall
normalization will not affect the location of the leading singular
support or the estimate of the typical energy scale. Here
\(\hat n\) and \(\hat n_0\) denote angular positions on \(S^{d-1}\).

We use the global AdS mode expansion
\begin{equation}
\Phi(\tau,\rho,\hat n)
=
\sum_{n,l,m}
\psi_{nl}(\rho)
\left[
 a_{nlm}e^{-i\omega_{nl}\tau}Y^*_{lm}(\hat n)
+a^\dagger_{nlm}e^{i\omega_{nl}\tau}Y_{lm}(\hat n)
\right],
\label{eq:app-bulk-mode-expansion}
\end{equation}
where
\begin{equation}
\omega_{nl}=2n+l+\Delta
\label{eq:app-global-frequency}
\end{equation}
is the dimensionless global energy, and
\begin{equation}
\psi_{nl}(\rho)
=
\frac{1}{\mathcal N_{nl}}
\sin^l\rho\,\cos^\Delta\rho\,
P_n^{\,l+d/2-1,\Delta-d/2}(\cos2\rho) .
\label{eq:app-radial-mode}
\end{equation}
The corresponding CFT primary on the cylinder is expanded in the same
creation and annihilation operators as
\begin{equation}
\Ocal_\Delta(\tau,\hat n)
=
\sum_{n,l,m}
\psi^{\rm CFT}_{nl}
\left[
Y_{lm}(\hat n)e^{i\omega_{nl}\tau}a^\dagger_{nlm}
+Y^*_{lm}(\hat n)e^{-i\omega_{nl}\tau}a_{nlm}
\right],
\label{eq:app-cft-mode-expansion}
\end{equation}
with
\begin{equation}
\psi^{\rm CFT}_{nl}
=
\left[
\frac{2}{\pi}
\frac{\Gamma(d/2)}
{\Gamma(\Delta)\Gamma(\Delta+1-d/2)}
\right]^{1/2}
\left[
\frac{\Gamma(n+\Delta+1-d/2)\Gamma(n+l+\Delta)}
{\Gamma(n+1)\Gamma(n+l+d/2)}
\right]^{1/2} .
\label{eq:app-cft-mode-coeff}
\end{equation}
These formulae follow from applying the extrapolate dictionary to
global AdS normal modes
\cite{Balasubramanian:1998sn,Banks:1998dd,Hamilton:2006az}.

Acting on the vacuum, the boundary operator creates the one-particle
state
\begin{equation}
\Ocal_\Delta(0,\hat n_0)|0\rangle
=
\sum_{n,l,m}
\psi^{\rm CFT}_{nl}Y_{lm}(\hat n_0)
a^\dagger_{nlm}|0\rangle .
\end{equation}
The Euclidean time regulator gives
\begin{equation}
e^{-\epsilon H}a^\dagger_{nlm}|0\rangle
=
e^{-\epsilon\omega_{nl}}a^\dagger_{nlm}|0\rangle .
\end{equation}
Only the annihilation part of the bulk field contributes in
\eqref{eq:app-heavy-profile-def}. Therefore
\begin{equation}
\Psi_\epsilon(\tau,\rho,\hat n;\hat n_0)
=
\sum_{n,l,m}
\psi_{nl}(\rho)\psi^{\rm CFT}_{nl}
e^{-(\epsilon+i\tau)\omega_{nl}}
Y^*_{lm}(\hat n)Y_{lm}(\hat n_0) .
\label{eq:app-heavy-mode-sum}
\end{equation}
This is the regulated bulk wave function written entirely in global
modes.

This mode sum is the global bulk-to-boundary Wightman function with the
boundary time shifted by \(-i\epsilon\). Up to an overall normalization
convention,
\begin{equation}
\Psi_\epsilon(\tau,\rho,\hat n;\hat n_0)
=
\mathcal C_\Delta
\left[
\frac{\cos\rho}
{2\left\{
\cos(\tau-i\epsilon)-\sin\rho\,\hat n\cdot\hat n_0
\right\}}
\right]^\Delta .
\label{eq:app-heavy-closed-form}
\end{equation}
For small \(\epsilon\),
\begin{equation}
\cos(\tau-i\epsilon)
=
\cos\tau+i\epsilon\sin\tau+O(\epsilon^2) .
\end{equation}
Thus, as \(\epsilon\to0\), the singular support of the regulated
profile approaches
\begin{equation}
\cos\tau=\sin\rho\,\hat n\cdot\hat n_0 .
\label{eq:app-null-cone}
\end{equation}
This is the bulk light cone emanating from the boundary insertion. In
the radial direction \(\hat n=\hat n_0\), it gives
\begin{equation}
\cos\tau=\sin\rho,
\qquad
\tau=\frac{\pi}{2}-\rho,
\end{equation}
which is the incoming radial null geodesic.

The singular-support argument can be sharpened by locating the maximum
of the regulated profile.  For large $\Delta$, the profile becomes
sharply localized around this maximum.  We define
\begin{equation}
x:=\sin\rho,
\qquad
\cos\gamma:=\hat n\cdot\hat n_0,
\end{equation}
where $0\leq x<1$ and $0\leq\gamma\leq\pi$.  Taking the absolute value
of \eqref{eq:app-heavy-closed-form} gives
\begin{equation}
\left|\Psi_\epsilon\right|^{2/\Delta}
\propto
\frac{1-x^2}{
\left(\cos\tau\cosh\epsilon-x\cos\gamma\right)^2
+\sin^2\tau\sinh^2\epsilon
}.
\label{eq:app-heavy-absolute-profile}
\end{equation}
For fixed $x$ and $\cos\tau\neq0$, maximizing over the angular
separation gives
\begin{equation}
\cos\gamma=
\begin{cases}
\operatorname{sgn}(\cos\tau),
& x\leq |\cos\tau|\cosh\epsilon,\\[2mm]
\displaystyle\frac{\cos\tau\cosh\epsilon}{x},
& x\geq |\cos\tau|\cosh\epsilon.
\end{cases}
\label{eq:app-heavy-angular-max}
\end{equation}
On the second branch, the denominator is independent of $x$, while the
numerator decreases with $x$.  The global maximum therefore lies on the
first branch.  Defining
\begin{equation}
A:=|\cos\tau|\cosh\epsilon,
\qquad
B^2:=\sin^2\tau\sinh^2\epsilon,
\end{equation}
the stationarity condition on this branch is
\begin{equation}
x\left[(A-x)^2+B^2\right]
-(1-x^2)(A-x)=0.
\end{equation}
The solution corresponding to the maximum in $0\leq x<1$ is
\begin{equation}
x_{\rm peak}(\tau)
=
\frac{|\cos\tau|}{\cosh\epsilon}.
\label{eq:app-heavy-xpeak}
\end{equation}
At $\cos\tau=0$, the maximum is at $x=0$, where the angular direction
is degenerate, consistently with \eqref{eq:app-heavy-xpeak}.

Since the maximum lies on the axis through the boundary insertion, it
is convenient to introduce the signed coordinate along this axis,
\begin{equation}
X:=x\cos\gamma
=
\sin\rho\,\hat n\cdot\hat n_0 .
\end{equation}
The peak locus is then
\begin{equation}
X_{\rm peak}(\tau)
=
\operatorname{sech}\epsilon\,\cos\tau
=
\cos\tau-\frac{\epsilon^2}{2}\cos\tau
+O(\epsilon^4).
\label{eq:app-heavy-peak-locus}
\end{equation}
In the same coordinate, the radial null geodesic emitted from the
boundary insertion at $\tau=0$ is
\begin{equation}
X_{\rm null}(\tau)=\cos\tau .
\end{equation}
Thus, in the regime $\epsilon\ll1$, the peak of the regulated bulk
profile approaches the radial null trajectory, with the first
finite-regulator displacement appearing at $O(\epsilon^2)$.  This
establishes the null approximation directly from the spatial profile,
in addition to the energy estimate below.

We now estimate the energy scale from the same mode sum. 
At the center of AdS, \(\rho=0\), only the \(l=0\) modes contribute
because of the factor \(\sin^l\rho\) in \eqref{eq:app-radial-mode}.
Hence
\begin{equation}
\Psi_\epsilon(\tau,0)
=
\sum_{n=0}^{\infty}
\psi_{n0}(0)\psi^{\rm CFT}_{n0}
e^{-(\epsilon+i\tau)(2n+\Delta)} .
\label{eq:app-center-profile}
\end{equation}
The \(n\)-dependence of the coefficient can be read off directly from
the closed form \eqref{eq:app-heavy-closed-form}. Setting
\(z=\tau-i\epsilon\), the profile at \(\rho=0\) is
\begin{equation}
\Psi_\epsilon(\tau,0)
\propto
\left(\frac{1}{2\cos z}\right)^\Delta
=
e^{-i\Delta z}
\left(1+e^{-2iz}\right)^{-\Delta}.
\end{equation}
Using
\begin{equation}
(1+x)^{-\Delta}
=
\sum_{n=0}^{\infty}
(-1)^n
\frac{\Gamma(n+\Delta)}
{\Gamma(\Delta)\Gamma(n+1)}
x^n ,
\end{equation}
we obtain
\begin{equation}
\Psi_\epsilon(\tau,0)
\propto
e^{-(\epsilon+i\tau)\Delta}
\sum_{n=0}^{\infty}
(-1)^n
\frac{\Gamma(n+\Delta)}{\Gamma(n+1)}
e^{-2n(\epsilon+i\tau)} .
\label{eq:app-center-mode-sum}
\end{equation}
Here the \(n\)-independent factor \(1/\Gamma(\Delta)\) has been
absorbed into the overall normalization.
For the magnitude of the mode weight,
\begin{equation}
w_n
\sim
\frac{\Gamma(n+\Delta)}{\Gamma(n+1)}e^{-2\epsilon n} .
\end{equation}
At large \(n\),
\begin{equation}
\frac{\Gamma(n+\Delta)}{\Gamma(n+1)}
\sim n^{\Delta-1} .
\end{equation}
Therefore the dominant contribution is determined by the saddle of
\begin{equation}
\log w_n
\simeq
(\Delta-1)\log n-2\epsilon n .
\end{equation}
The saddle condition gives
\begin{equation}
\frac{\Delta-1}{n}-2\epsilon=0,
\qquad
n_{\rm typ}
\simeq
\frac{\Delta}{2\epsilon} .
\end{equation}
The corresponding dimensionless global energy is
\begin{equation}
\omega_{\rm typ}
=
2n_{\rm typ}+\Delta
\sim
\frac{\Delta}{\epsilon} .
\label{eq:app-typical-energy}
\end{equation}
This estimate was already obtained in \cite{Berenstein:2019qmm}, and is
also consistent with the CFT result \eqref{eq:LCD-energy}.
This is not a dimensionful energy. Restoring the AdS radius,
\begin{equation}
E_{\rm typ}^{\rm phys}
=
\frac{\omega_{\rm typ}}{L}
\sim
\frac{\Delta}{L\epsilon} .
\end{equation}
For a scalar field dual to a scalar primary operator of conformal dimension \(\Delta\),
\begin{equation}
M^2L^2=\Delta(\Delta-d),
\qquad
ML\simeq\Delta.
\end{equation}
Thus \(ML\) is the dimensionless rest energy of
the dual bulk excitation, while \(M\) is its physical rest energy.
Hence
\begin{equation}
\frac{E_{\rm typ}^{\rm phys}}{M}
=
\frac{\omega_{\rm typ}}{ML}
\sim
\frac{1}{\epsilon} .
\end{equation}
In the regime \(\epsilon\ll1\), the LCD state has a dominant
energy parametrically larger than \(ML\), consistently with the direct
approach of its peak locus to the radial null trajectory in
\eqref{eq:app-heavy-peak-locus}. At the same time, the more precise
condition for neglecting gravitational backreaction is not simply
\(\Delta\ll c\), but rather
\begin{equation}
\omega_{\rm typ}\sim \frac{\Delta}{\epsilon}\ll c .
\end{equation}
This is because the strength of the backreaction is controlled by the
dimensionless energy \(E_{\rm typ}^{\rm phys}L\simeq\omega_{\rm typ}\). Under
this probe-limit condition, the leading bulk propagation is governed by
the radial null limit of the semiclassical excitation.


\begin{thebibliography}{10}

\bibitem{Maldacena:1997re}
J.~M. Maldacena, ``{The Large N limit of superconformal field theories and supergravity},'' \href{https://dx.doi.org/10.4310/ATMP.1998.v2.n2.a1}{{\em Adv. Theor. Math. Phys.} {\bfseries 2} (1998) 231--252}, \href{https://arxiv.org/abs/hep-th/9711200}{{\ttfamily arXiv:hep-th/9711200}}.

\bibitem{Gubser:1998bc}
S.~S. Gubser, I.~R. Klebanov, and A.~M. Polyakov, ``{Gauge theory correlators from noncritical string theory},'' \href{https://dx.doi.org/10.1016/S0370-2693(98)00377-3}{{\em Phys. Lett. B} {\bfseries 428} (1998) 105--114}, \href{https://arxiv.org/abs/hep-th/9802109}{{\ttfamily arXiv:hep-th/9802109}}.

\bibitem{Witten:1998qj}
E.~Witten, ``{Anti-de Sitter space and holography},'' \href{https://dx.doi.org/10.4310/ATMP.1998.v2.n2.a2}{{\em Adv. Theor. Math. Phys.} {\bfseries 2} (1998) 253--291}, \href{https://arxiv.org/abs/hep-th/9802150}{{\ttfamily arXiv:hep-th/9802150}}.

\bibitem{Hamilton:2006fh}
A.~Hamilton, D.~N. Kabat, G.~Lifschytz, and D.~A. Lowe, ``{Local bulk operators in AdS/CFT: A boundary view of horizons and locality},'' \href{https://dx.doi.org/10.1103/PhysRevD.73.086003}{{\em Phys. Rev. D} {\bfseries 73} (2006) 086003}, \href{https://arxiv.org/abs/hep-th/0506118}{{\ttfamily arXiv:hep-th/0506118}}.

\bibitem{Hamilton:2006az}
A.~Hamilton, D.~N. Kabat, G.~Lifschytz, and D.~A. Lowe, ``{Holographic representation of local bulk operators},'' \href{https://dx.doi.org/10.1103/PhysRevD.74.066009}{{\em Phys. Rev. D} {\bfseries 74} (2006) 066009}, \href{https://arxiv.org/abs/hep-th/0606141}{{\ttfamily arXiv:hep-th/0606141}}.

\bibitem{Kabat:2011rz}
D.~Kabat, G.~Lifschytz, and D.~A. Lowe, ``{Constructing local bulk observables in interacting AdS/CFT},'' \href{https://dx.doi.org/10.1103/PhysRevD.83.106009}{{\em Phys. Rev. D} {\bfseries 83} (2011) 106009}, \href{https://arxiv.org/abs/1102.2910}{{\ttfamily arXiv:1102.2910 [hep-th]}}.

\bibitem{Harlow:2018fse}
D.~Harlow, ``{TASI Lectures on the Emergence of Bulk Physics in AdS/CFT},'' \href{https://dx.doi.org/10.22323/1.305.0002}{{\em PoS} {\bfseries TASI2017} (2018) 002}, \href{https://arxiv.org/abs/1802.01040}{{\ttfamily arXiv:1802.01040 [hep-th]}}.

\bibitem{Hubeny:2006yu}
V.~E. Hubeny, H.~Liu, and M.~Rangamani, ``{Bulk-cone singularities and signatures of horizon formation in AdS/CFT},'' \href{https://dx.doi.org/10.1088/1126-6708/2007/01/009}{{\em JHEP} {\bfseries 01} (2007) 009}, \href{https://arxiv.org/abs/hep-th/0610041}{{\ttfamily arXiv:hep-th/0610041}}.

\bibitem{Dodelson:2023nnr}
M.~Dodelson, C.~Iossa, R.~Karlsson, A.~Lupsasca, and A.~Zhiboedov, ``{Black hole bulk-cone singularities},'' \href{https://dx.doi.org/10.1007/JHEP07(2024)046}{{\em JHEP} {\bfseries 07} (2024) 046}, \href{https://arxiv.org/abs/2310.15236}{{\ttfamily arXiv:2310.15236 [hep-th]}}.

\bibitem{Tanahashi:2025fqi}
N.~Tanahashi, S.~Terashima, and S.~Yoshikawa, ``{Holographic description of bulk wave packets in AdS$_{4}$/CFT$_{3}$},'' \href{https://dx.doi.org/10.1007/JHEP06(2025)214}{{\em JHEP} {\bfseries 06} (2025) 214}, \href{https://arxiv.org/abs/2503.11485}{{\ttfamily arXiv:2503.11485 [hep-th]}}.

\bibitem{Terashima:2020tub}
S.~Terashima, ``{Bulk locality in the AdS/CFT correspondence},'' \href{https://dx.doi.org/10.1103/PhysRevD.104.086014}{{\em Phys. Rev. D} {\bfseries 104} no.~8, (2021) 086014}, \href{https://arxiv.org/abs/2005.05962}{{\ttfamily arXiv:2005.05962 [hep-th]}}.

\bibitem{Terashima:2023mcr}
S.~Terashima, ``{Wave packets in AdS/CFT correspondence},'' \href{https://dx.doi.org/10.1103/PhysRevD.109.106012}{{\em Phys. Rev. D} {\bfseries 109} no.~10, (2024) 106012}, \href{https://arxiv.org/abs/2304.08478}{{\ttfamily arXiv:2304.08478 [hep-th]}}.

\bibitem{Gary:2009ae}
M.~Gary, S.~B. Giddings, and J.~Penedones, ``{Local bulk S-matrix elements and CFT singularities},'' \href{https://dx.doi.org/10.1103/PhysRevD.80.085005}{{\em Phys. Rev. D} {\bfseries 80} (2009) 085005}, \href{https://arxiv.org/abs/0903.4437}{{\ttfamily arXiv:0903.4437 [hep-th]}}.

\bibitem{Gary:2009mi}
M.~Gary and S.~B. Giddings, ``{The Flat space S-matrix from the AdS/CFT correspondence?},'' \href{https://dx.doi.org/10.1103/PhysRevD.80.046008}{{\em Phys. Rev. D} {\bfseries 80} (2009) 046008}, \href{https://arxiv.org/abs/0904.3544}{{\ttfamily arXiv:0904.3544 [hep-th]}}.

\bibitem{Nozaki:2013wia}
M.~Nozaki, T.~Numasawa, and T.~Takayanagi, ``{Holographic Local Quenches and Entanglement Density},'' \href{https://dx.doi.org/10.1007/JHEP05(2013)080}{{\em JHEP} {\bfseries 05} (2013) 080}, \href{https://arxiv.org/abs/1302.5703}{{\ttfamily arXiv:1302.5703 [hep-th]}}.

\bibitem{Caputa:2014vaa}
P.~Caputa, M.~Nozaki, and T.~Takayanagi, ``{Entanglement of local operators in large-N conformal field theories},'' \href{https://dx.doi.org/10.1093/ptep/ptu122}{{\em PTEP} {\bfseries 2014} (2014) 093B06}, \href{https://arxiv.org/abs/1405.5946}{{\ttfamily arXiv:1405.5946 [hep-th]}}.

\bibitem{Berenstein:2019qmm}
D.~Berenstein and J.~Simon, ``{Localized states in global AdS space},'' \href{https://dx.doi.org/10.1103/PhysRevD.101.046026}{{\em Phys. Rev. D} {\bfseries 101} no.~4, (2020) 046026}, \href{https://arxiv.org/abs/1910.10227}{{\ttfamily arXiv:1910.10227 [hep-th]}}.

\bibitem{Marolf:2017kvq}
D.~Marolf, O.~Parrikar, C.~Rabideau, A.~Izadi~Rad, and M.~Van~Raamsdonk, ``{From Euclidean Sources to Lorentzian Spacetimes in Holographic Conformal Field Theories},'' \href{https://dx.doi.org/10.1007/JHEP06(2018)077}{{\em JHEP} {\bfseries 06} (2018) 077}, \href{https://arxiv.org/abs/1709.10101}{{\ttfamily arXiv:1709.10101 [hep-th]}}.

\bibitem{Jia:2026pmv}
Y.~Jia and M.~Kulaxizi, ``{Bulk Phase Shift and Singularity},'' \href{https://arxiv.org/abs/2602.06558}{{\ttfamily arXiv:2602.06558 [hep-th]}}.

\bibitem{duffin_lecture}
D.~Simmons-Duffin, ``Tasi lectures on conformal field theory in lorentzian signature,''
\newblock \url{https://www.desy.de/~bargheer/string-journal-club/presentations/2023-05-16_Sebastian-Harris_Simmons-Duffin:_TASI-Lorentzian-CFT.pdf}.

\bibitem{Gao:2000ga}
S.~Gao and R.~M. Wald, ``{Theorems on gravitational time delay and related issues},'' \href{https://dx.doi.org/10.1088/0264-9381/17/24/305}{{\em Class. Quant. Grav.} {\bfseries 17} (2000) 4999--5008}, \href{https://arxiv.org/abs/gr-qc/0007021}{{\ttfamily arXiv:gr-qc/0007021}}.

\bibitem{Hofman:2008ar}
D.~M. Hofman and J.~Maldacena, ``Conformal collider physics: Energy and charge correlations,'' \href{https://dx.doi.org/10.1088/1126-6708/2008/05/012}{{\em JHEP} {\bfseries 05} (2008) 012}, \href{https://arxiv.org/abs/0803.1467}{{\ttfamily arXiv:0803.1467 [hep-th]}}.

\bibitem{Kravchuk:2018htv}
P.~Kravchuk and D.~Simmons-Duffin, ``Light-ray operators in conformal field theory,'' \href{https://dx.doi.org/10.1007/JHEP11(2018)102}{{\em JHEP} {\bfseries 11} (2018) 102}, \href{https://arxiv.org/abs/1805.00098}{{\ttfamily arXiv:1805.00098 [hep-th]}}.

\bibitem{Faulkner:2016mzt}
T.~Faulkner, R.~G. Leigh, O.~Parrikar, and H.~Wang, ``Modular hamiltonians for deformed half-spaces and the averaged null energy condition,'' \href{https://dx.doi.org/10.1007/JHEP09(2016)038}{{\em JHEP} {\bfseries 09} (2016) 038}, \href{https://arxiv.org/abs/1605.08072}{{\ttfamily arXiv:1605.08072 [hep-th]}}.

\bibitem{Chen:2025jbf}
H.-Y. Chen, Y.~Hikida, and Y.~Koga, ``{Bulk-cone singularities and echoes from AdS exotic compact objects},'' \href{https://dx.doi.org/10.1007/JHEP04(2026)210}{{\em JHEP} {\bfseries 04} (2026) 210}, \href{https://arxiv.org/abs/2512.21535}{{\ttfamily arXiv:2512.21535 [hep-th]}}.

\bibitem{Bousso:2012mh}
R.~Bousso, B.~Freivogel, S.~Leichenauer, V.~Rosenhaus, and C.~Zukowski, ``{Null Geodesics, Local CFT Operators and AdS/CFT for Subregions},'' \href{https://dx.doi.org/10.1103/PhysRevD.88.064057}{{\em Phys. Rev. D} {\bfseries 88} (2013) 064057}, \href{https://arxiv.org/abs/1209.4641}{{\ttfamily arXiv:1209.4641 [hep-th]}}.

\bibitem{Maldacena:2015iua}
J.~Maldacena, D.~Simmons-Duffin, and A.~Zhiboedov, ``{Looking for a bulk point},'' {\em JHEP} {\bfseries 01} (2017) 013, \href{https://arxiv.org/abs/1509.03612}{{\ttfamily arXiv:1509.03612 [hep-th]}}.

\bibitem{Kinoshita:2023lgy}
S.~Kinoshita, K.~Murata, and D.~Takeda, ``{Shooting null geodesics into holographic spacetimes},'' \href{https://dx.doi.org/10.1007/JHEP10(2023)074}{{\em JHEP} {\bfseries 10} (2023) 074}, \href{https://arxiv.org/abs/2304.01936}{{\ttfamily arXiv:2304.01936 [hep-th]}}.

\bibitem{Tanahashi:2026mia}
N.~Tanahashi, S.~Terashima, and S.~Yoshikawa, ``{Probing Black Hole Thermal Effects in the Dual CFT via Wave Packets},'' \href{https://arxiv.org/abs/2601.04647}{{\ttfamily arXiv:2601.04647 [hep-th]}}.

\bibitem{Balasubramanian:1999re}
V.~Balasubramanian and P.~Kraus, ``{A Stress tensor for Anti-de Sitter gravity},'' \href{https://dx.doi.org/10.1007/s002200050764}{{\em Commun. Math. Phys.} {\bfseries 208} (1999) 413--428}, \href{https://arxiv.org/abs/hep-th/9902121}{{\ttfamily arXiv:hep-th/9902121}}.

\bibitem{deHaro:2000vlm}
S.~de~Haro, K.~Skenderis, and S.~N. Solodukhin, ``{Holographic reconstruction of space-time and renormalization in the AdS/CFT correspondence},'' \href{https://dx.doi.org/10.1007/s002200100381}{{\em Commun. Math. Phys.} {\bfseries 217} (2001) 595--622}, \href{https://arxiv.org/abs/hep-th/0002230}{{\ttfamily arXiv:hep-th/0002230}}.

\bibitem{Bantilan:2012vu}
H.~Bantilan, F.~Pretorius, and S.~S. Gubser, ``{Simulation of Asymptotically AdS5 Spacetimes with a Generalized Harmonic Evolution Scheme},'' \href{https://dx.doi.org/10.1103/PhysRevD.85.084038}{{\em Phys. Rev. D} {\bfseries 85} (2012) 084038}, \href{https://arxiv.org/abs/1201.2132}{{\ttfamily arXiv:1201.2132 [hep-th]}}.

\bibitem{deBoer:2026cng}
J.~de~Boer, J.~Hollander, and A.~Rolph, ``{A Reverse Black Hole Information Problem},'' \href{https://arxiv.org/abs/2601.22077}{{\ttfamily arXiv:2601.22077 [hep-th]}}.

\bibitem{Balasubramanian:1998sn}
V.~Balasubramanian, P.~Kraus, and A.~E. Lawrence, ``{Bulk versus boundary dynamics in anti-de Sitter space-time},'' \href{https://dx.doi.org/10.1103/PhysRevD.59.046003}{{\em Phys. Rev. D} {\bfseries 59} (1999) 046003}, \href{https://arxiv.org/abs/hep-th/9805171}{{\ttfamily arXiv:hep-th/9805171}}.

\bibitem{Banks:1998dd}
T.~Banks, M.~R. Douglas, G.~T. Horowitz, and E.~J. Martinec, ``{AdS dynamics from conformal field theory},'' \href{https://arxiv.org/abs/hep-th/9808016}{{\ttfamily arXiv:hep-th/9808016}}.

\bibitem{Shibuya:2025rel}
S.~Shibuya and S.~Sugishita, ``{Notes on Rindler wave packets in Minkowski spacetime},'' \href{https://dx.doi.org/10.1103/dfzz-jnsh}{{\em Phys. Rev. D} {\bfseries 112} no.~12, (2025) 125007}, \href{https://arxiv.org/abs/2505.20078}{{\ttfamily arXiv:2505.20078 [hep-th]}}.

\end{thebibliography}
\providecommand{\href}[2]{#2}\begingroup\raggedright\endgroup

\end{document}